\documentclass[%
 superscriptaddress,
 preprint,
 notitlepage,
 showpacs,
 amsmath,
 amssymb,
 aps,
 prc
]{revtex4-1}

\usepackage{graphicx}% Include figure files
\usepackage{dcolumn}% Align table columns on decimal point
\usepackage{bm}% bold math
\begin{document}

\title{Photoproduction of $\Lambda$ and $\Sigma^0$ hyperons using linearly polarized photons}

% === Author Section ======================================

%%%%%%%%%%%%%% Lead author institution goes first     %%%%%%%%%%%%%%%%%%%%%%%%

\newcommand*{\GLASGOW}{University of Glasgow, Glasgow G12 8QQ, United Kingdom}
\newcommand*{\GLASGOWindex}{1}
\affiliation{\GLASGOW}

%%%%%%%%%%%%%%% Latex Macros for institute addresses  %%%%%%%%%%%%%%%%%%%%%%%%% 

\newcommand*{\ANL}{Argonne National Laboratory, Argonne, Illinois 60439}
\newcommand*{\ANLindex}{1}
\affiliation{\ANL}
\newcommand*{\ASU}{Arizona State University, Tempe, Arizona 85287-1504}
\newcommand*{\ASUindex}{2}
\affiliation{\ASU}
\newcommand*{\CSUDH}{California State University, Dominguez Hills, Carson, CA 90747}
\newcommand*{\CSUDHindex}{3}
\affiliation{\CSUDH}
\newcommand*{\CMU}{Carnegie Mellon University, Pittsburgh, Pennsylvania 15213}
\newcommand*{\CMUindex}{4}
\affiliation{\CMU}
\newcommand*{\CUA}{Catholic University of America, Washington, D.C. 20064}
\newcommand*{\CUAindex}{5}
\affiliation{\CUA}
\newcommand*{\SACLAY}{CEA, Centre de Saclay, Irfu/Service de Physique Nucl\'eaire, 91191 Gif-sur-Yvette, France}
\newcommand*{\SACLAYindex}{6}
\affiliation{\SACLAY}
\newcommand*{\UCONN}{University of Connecticut, Storrs, Connecticut 06269}
\newcommand*{\UCONNindex}{7}
\affiliation{\UCONN}
\newcommand*{\EDINBURGH}{University of Edinburgh, Edinburgh EH9 3JZ, United Kingdom}
\newcommand*{\EDINBURGHindex}{7}
\affiliation{\EDINBURGH}
\newcommand*{\FU}{Fairfield University, Fairfield, CT 06824}
\newcommand*{\FUindex}{8}
\affiliation{\FU}
\newcommand*{\FIU}{Florida International University, Miami, Florida 33199}
\newcommand*{\FIUindex}{8}
\affiliation{\FIU}
\newcommand*{\FSU}{Florida State University, Tallahassee, Florida 32306}
\newcommand*{\FSUindex}{9}
\affiliation{\FSU}
\newcommand*{\Genova}{Universit$\grave{a}$ di Genova, 16146 Genova, Italy}
\newcommand*{\Genovaindex}{10}
\affiliation{\Genova}
\newcommand*{\GWUI}{The George Washington University, Washington, DC 20052}
\newcommand*{\GWUIindex}{11}
\affiliation{\GWUI}
\newcommand*{\ISU}{Idaho State University, Pocatello, Idaho 83209}
\newcommand*{\ISUindex}{12}
\affiliation{\ISU}
\newcommand*{\INFNFE}{INFN, Sezione di Ferrara, 44100 Ferrara, Italy}
\newcommand*{\INFNFEindex}{13}
\affiliation{\INFNFE}
\newcommand*{\INFNFR}{INFN, Laboratori Nazionali di Frascati, 00044 Frascati, Italy}
\newcommand*{\INFNFRindex}{14}
\affiliation{\INFNFR}
\newcommand*{\INFNGE}{INFN, Sezione di Genova, 16146 Genova, Italy}
\newcommand*{\INFNGEindex}{15}
\affiliation{\INFNGE}
\newcommand*{\INFNRO}{INFN, Sezione di Roma Tor Vergata, 00133 Rome, Italy}
\newcommand*{\INFNROindex}{16}
\affiliation{\INFNRO}
\newcommand*{\INFNTUR}{INFN, Sezione di Torino, 10125 Torino, Italy}
\newcommand*{\INFNTURindex}{17}
\affiliation{\INFNTUR}
\newcommand*{\ORSAY}{Institut de Physique Nucl\'eaire, CNRS/IN2P3 and Universit\'e Paris Sud, Orsay, France}
\newcommand*{\ORSAYindex}{18}
\affiliation{\ORSAY}
\newcommand*{\ITEP}{Institute of Theoretical and Experimental Physics, Moscow, 117259, Russia}
\newcommand*{\ITEPindex}{19}
\affiliation{\ITEP}
\newcommand*{\JMU}{James Madison University, Harrisonburg, Virginia 22807}
\newcommand*{\JMUindex}{20}
\affiliation{\JMU}
\newcommand*{\KNU}{Kyungpook National University, Daegu 702-701, Republic of Korea}
\newcommand*{\KNUindex}{21}
\affiliation{\KNU}
\newcommand*{\MISS}{Mississippi State University, Mississippi State, MS 39762-5167}
\newcommand*{\MISSindex}{22}
\affiliation{\MISS}
\newcommand*{\UNH}{University of New Hampshire, Durham, New Hampshire 03824-3568}
\newcommand*{\UNHindex}{23}
\affiliation{\UNH}
\newcommand*{\NSU}{Norfolk State University, Norfolk, Virginia 23504}
\newcommand*{\NSUindex}{25}
\affiliation{\NSU}
\newcommand*{\OHIOU}{Ohio University, Athens, Ohio  45701}
\newcommand*{\OHIOUindex}{26}
\affiliation{\OHIOU}
\newcommand*{\ODU}{Old Dominion University, Norfolk, Virginia 23529}
\newcommand*{\ODUindex}{27}
\affiliation{\ODU}
\newcommand*{\RPI}{Rensselaer Polytechnic Institute, Troy, New York 12180-3590}
\newcommand*{\RPIindex}{28}
\affiliation{\RPI}
\newcommand*{\URICH}{University of Richmond, Richmond, Virginia 23173}
\newcommand*{\URICHindex}{29}
\affiliation{\URICH}
\newcommand*{\ROMAII}{Universita' di Roma Tor Vergata, 00133 Rome Italy}
\newcommand*{\ROMAIIindex}{30}
\affiliation{\ROMAII}
\newcommand*{\MSU}{Skobeltsyn Institute of Nuclear Physics, Lomonosov Moscow State University, 119234 Moscow, Russia}
\newcommand*{\MSUindex}{31}
\affiliation{\MSU}
\newcommand*{\SCAROLINA}{University of South Carolina, Columbia, South Carolina 29208}
\newcommand*{\SCAROLINAindex}{32}
\affiliation{\SCAROLINA}
\newcommand*{\TEMPLE}{Temple University,  Philadelphia, PA 19122 }
\newcommand*{\TEMPLEindex}{33}
\affiliation{\TEMPLE}
\newcommand*{\JLAB}{Thomas Jefferson National Accelerator Facility, Newport News, Virginia 23606}
\newcommand*{\JLABindex}{34}
\affiliation{\JLAB}
\newcommand*{\UTFSM}{Universidad T\'{e}cnica Federico Santa Mar\'{i}a, Casilla 110-V Valpara\'{i}so, Chile}
\newcommand*{\UTFSMindex}{35}
\affiliation{\UTFSM}
\newcommand*{\VT}{Virginia Tech, Blacksburg, Virginia   24061-0435}
\newcommand*{\VTindex}{37}
\affiliation{\VT}
\newcommand*{\VIRGINIA}{University of Virginia, Charlottesville, Virginia 22901}
\newcommand*{\VIRGINIAindex}{38}
\affiliation{\VIRGINIA}
\newcommand*{\WM}{College of William and Mary, Williamsburg, Virginia 23187-8795}
\newcommand*{\WMindex}{39}
\affiliation{\WM}
\newcommand*{\YEREVAN}{Yerevan Physics Institute, 375036 Yerevan, Armenia}
\newcommand*{\YEREVANindex}{40}
\affiliation{\YEREVAN}

\newcommand*{\NOWJLAB}{Thomas Jefferson National Accelerator Facility, Newport News, Virginia 23606}

\newcommand*{\NOWGLASGOW}{University of Glasgow, Glasgow G12 8QQ, United Kingdom}

\newcommand*{\NOWINFNGE}{INFN, Sezione di Genova, 16146 Genova, Italy}
 %%%%%%%%%%%%%%% END OF Latex Macros for institute addresses  %%%%%%%%%%%%%%%%

%%%%%%%%%%%%%% Lead authors     %%%%%%%%%%%%%%%%%%%%%%%%%%%%%%%%%%%%%%%%%%%%%%
\author{C.A.~Paterson}
\altaffiliation[Current address: ]{Nuclear Cardiology and PET Centre, NHS Glasgow}
\affiliation{\GLASGOW}

\author{D.G.~Ireland}
\email[Corresponding author: ]{David.Ireland@glasgow.ac.uk}
\affiliation{\GLASGOW}

\author{K.~Livingston}
\affiliation{\GLASGOW}

\author{B.~McKinnon}
\affiliation{\GLASGOW}

%%%%%%%%%%%%%% Other authors    %%%%%%%%%%%%%%%%%%%%%%%%%%%%%%%%%%%%%%%%%%%%%%

\author {K.P. ~Adhikari} 
\affiliation{\MISS}
\author {D.~Adikaram} 
\affiliation{\JLAB}
\affiliation{\ODU}
\author {Z.~Akbar} 
\affiliation{\FSU}
\author {M.~Amaryan} 
\affiliation{\ODU}
\author {S. ~Anefalos~Pereira} 
\affiliation{\INFNFR}
\author {R.A.~Badui} 
\affiliation{\FIU}
% ...

%%%%%%%%%%%%%% END of other authors %%%%%%%%%%%%%%%%%%%%%%%%%%%%%%%%%%%%%%%%%%

\collaboration{CLAS Collaboration}
\noaffiliation

% === Abstract ============================================

\date{\today}

\begin{abstract}
\begin{description}
\item[Background]
Measurements of polarization observables for the reactions $\vec{\gamma} p \rightarrow K^+ \Lambda$ and $\vec{\gamma} p \rightarrow K^+ \Sigma^0$ have been performed. This is part of a programme of measurements designed to study the spectrum of baryon resonances in particular, and non-perturbative QCD in general.
\item[Purpose]
The accurate measurement of several polarization observables provides 
tight constraints for phenomenological fits, which allow the study of strangeness in nucleon and nuclear systems. Beam-recoil observables for the $\vec{\gamma} p \rightarrow K^+ \Sigma^0$ reaction have not been reported before now.
\item[Method]
The measurements were carried out using linearly polarized photon beams incident on a liquid hydrogen target, and the CLAS detector at the Thomas Jefferson National Accelerator Facility. The energy range of the results is 1.71\,GeV $<W<$ 2.19\,GeV, with an angular range $-0.75 < \cos\theta^{\star}_K < +0.85$.
\item[Results]
The observables extracted for both reactions are beam asymmetry $\Sigma$, target asymmetry $T$, and the beam-recoil double polarization observables $O_x$ and $O_z$. 
\item[Conclusions]
Comparison with theoretical fits indicates that in the regions where no previous data existed, the new data contain significant new information, and strengthen the evidence for the set of resonances used in the latest Bonn-Gatchina fit.
\end{description}
\end{abstract}

% PACS, the Physics and Astronomy Classification Scheme. 
% Still required? AIP Thesaurus?

%\keywords{Suggested keywords}%Use showkeys class option if keyword
                              %display desired
\maketitle

% === Introduction ========================================

\section{\label{sec:introduction}Introduction}

A critical ingredient in the understanding of QCD in the non-perturbative regime is a detailed knowledge of the spectrum of hadrons. In addition to being able to describe the nature of resonant states, one must also establish what resonant states do actually exist.

In the baryon sector, the quark model has provided useful guidance on which resonances to expect \cite{capstick_quark_2000}, and the general pattern and number of states have recently been by-and-large confirmed by lattice QCD results \cite{edwards_excited_2011}. A common feature of these predictions is that there are more predicted than observed resonances, which has led to the notion of \emph{missing resonances}. 

Most of the information about the spectrum of $N^{\star}$s and $\Delta^{\star}$s was derived from $\pi N$ scattering reactions, and indeed in 1983 it was thought by some that there was no realistic prospect of obtaining more information \cite{hey_baryon_1983}. However, the development of new experimental facilities and techniques has provided measurements sensitive to baryon resonances, particularly through photo- and electro-production of mesons. The number of measured states is slowly increasing  \cite{olive_review_2014}, but many predicted states remain unobserved. The current situation is summarized in \cite{crede_progress_2013}. 

Photoproduction of kaons, with associated $\Lambda$ and $\Sigma^0$ hyperons, is worthy of investigation. It is quite possible that through the strange decays of nonstrange baryons, some resonances may reveal themselves, when they would otherwise remain hidden in other channels \cite{capstick_strange_1998}. Another advantage of such reactions is that in the decays of the ground state $\Lambda$, its polarization is accessible due to its self-analyzing weak decay, where the degree of polarization can be measured from the angular distribution of the decay products. 

Pseudoscalar meson photoproduction is described by four complex amplitudes. Up to an overall phase, these amplitudes as functions of hadronic mass $W$ and center of mass meson scattering angle $\theta^{\star}$ (or Mandelstam variables $s$ and $t$) encode everything about the reaction, including the effects of any participating resonances, and so their extraction is an important goal. Such an extraction requires the measurement of a well chosen set of polarization observables \cite{chiang_completeness_1997} (for mathematical completeness) to an adequate level of accuracy \cite{ireland_information_2010}

A comprehensive set of measurements of differential cross sections, recoil polarizations and beam-recoil double polarisations, $C_x$ and $C_z$, for the reactions $\vec{\gamma} p \rightarrow K^+ \Lambda$ and $\vec{\gamma} p \rightarrow K^+ \Sigma^0$ has been carried out using the CEBAF Large Acceptance Spectrometer (CLAS) at Jefferson Lab \cite{mcnabb_hyperon_2004,bradford_differential_2006,bradford_first_2007,mccracken_differential_2010,dey_differential_2010}. Measurements of the beam asymmetry $\Sigma$ observable in these reactions have been reported by the LEPS \cite{zegers_beam-polarization_2003} and GRAAL \cite{lleres_polarization_2007} collaborations. The GRAAL collaboration also measured target asymmetry $T$, and the beam-recoil double polarization observables $O_x$ and $O_z$ for the $\vec{\gamma} p \rightarrow K^+ \Lambda$ reaction only \cite{lleres_measurement_2009}. 

In this article, we report measurements of the observables $\Sigma$, $T$, $O_x$ and $O_z$ for the reactions $\vec{\gamma} p \rightarrow K^+ \Lambda$ and $\vec{\gamma} p \rightarrow K^+ \Sigma^0$ in the energy range 1.71\,GeV $<W<$ 2.19\,GeV, and the angular range $-0.75 < \cos\theta^{\star}_K < +0.85$ \footnote{These measurements are also sensitive to the recoil polarization $P$, but since the measurements of $P$ reported in \citep{mccracken_differential_2010,dey_differential_2010} were of greater accuracy and covered a larger range in $W$, we chose to \emph{use} those results in the extraction procedure, having first established that the values of $P$ that could be measured in the present experiment were consistent with the previous ones.}, where $\theta^{\star}_K$ is the center of mass kaon scattering angle. The range in $W$ and $\cos\theta^{\star}_K$ covered in this measurement overlaps and extends the regions covered in the previous measurements. The results in the regions where the current experiment has overlaps with LEPS and GRAAL have significantly improved statistical accuracy for all measured observables, and the measurements of $T$, $O_x$ and $O_z$ for the $\vec{\gamma} p \rightarrow K^+ \Sigma^0$ reaction represent an entirely new data set.

% === Experimental Setup ==================================

\section{\label{sec:experiment}Experimental Setup}

The Thomas Jefferson National Accelerator Facility (JLab) in Newport News, Virginia is the site of the Continuous Electron Beam Accelerator Facility (CEBAF), which prior to its energy upgrade delivered beams of electrons of up to 6\,GeV. Beams of linearly polarized photons were produced using the coherent bremsstrahlung technique \cite{timm_coherent_1969,lohmann_linearly_1994}, which involves scattering electrons from a diamond radiator and detecting them in a tagging spectrometer \cite{sober_bremsstrahlung_2000}. The results reported here are part of a set of measurements known as the g8 run period, which were the first experiments to use linearly polarized photon beams with CLAS. 

The experimental configuration used for g8 consisted of a 4.55\,GeV electron beam incident on a 50\,$\mu$m thick diamond radiator. The polarization orientation of the photon beam was controlled by the careful alignment of the diamond radiator \cite{livingston_stonehenge_2009}. The diamond was mounted in a goniometer, and by orienting it at different angles, the photon energy at which the degree of polarization is at a maximum (known as the ``coherent edge'') could be varied. Coherent edge settings at 1.3, 1.5, 1.7, 1.9 and 2.1\,GeV were used in this run period. The degree of photon polarization was determined via a fit with a QED calculation \cite{livingston_polarization_2011}.

\begin{figure}
\includegraphics[width=0.58\columnwidth]{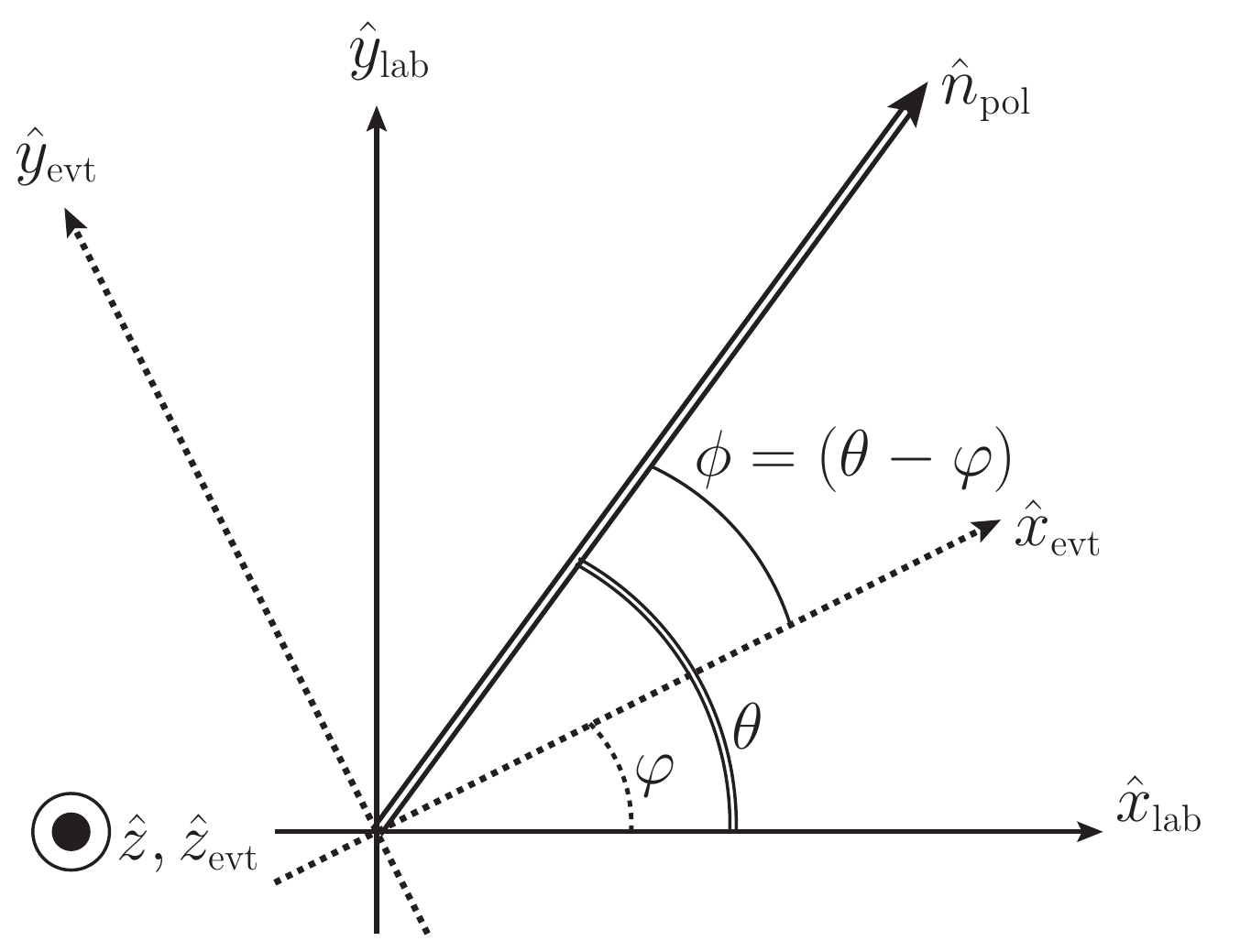}\protect\caption{\label{fig:axes}(Taken from \cite{dey_polarization_2011}) The definitions of lab and event axes, as well as azimuthal angles. The common lab, center-of-mass and event $z$-axis is directed out of the page. The lab $x$- and $y$-axes are in the horizontal and vertical directions, and the event $y$-axis is normal to the reaction plane.}
\end{figure}

Figure \ref{fig:axes} shows the general definition of directions. The lab axes $\hat{x}_{\mathrm{lab}},\hat{y}_{\mathrm{lab}}$ refer to the horizontal and vertical directions of the detector system. The coordinate system employed in this analysis is the so-called ``unprimed'' frame where, for a photon momentum $\vec{k}$ and a kaon momentum $\vec{q}$, axes are defined such that
\[
\hat{z}_{evt}=\frac{\vec{k}}{|\vec{k}|};\quad\hat{y}_{evt}=\frac{\vec{k}\times\vec{q}}{|\vec{k}\times\vec{q}|};\quad\hat{x}_{evt}=\hat{y}_{evt}\times\hat{z}_{evt}.
\]
In Fig.\,\ref{fig:axes}, therefore, $\vec{q}$ lies in the plane spanned by the vectors $\hat{x}_{evt}$ and $\hat{z}_{evt}$. The azimuthal angle $\phi$ is related to the measured azimuthal angle of the event $\varphi$ and the orientation of the polarization of the photon $\theta$ by:
\[
\phi = \theta - \varphi.
\]
In addition to varying the coherent edge setting, the orientation of the photon polarization axis could be controlled. The direction of photon polarization $\hat{n}_{\mathrm{pol}}$ was set by the goniometer orientation, and is defined relative to the lab axes. 

In practice, two settings of the orientation of photon polarization are employed: parallel (labelled $\parallel$), where the polarization axis is in the plane of the floor of the experimental hall ($\hat{x}_{\mathrm{lab}}$); perpendicular (labelled $\perp$), where it is oriented vertically ($\hat{y}_{\mathrm{lab}}$). Using these two settings, it is possible to form asymmetries in the measurements and extract several polarization observables. During the run the setting was switched from parallel to perpendicular, to accumulate similar numbers of events in each setting. Some runs were also taken where electrons were incident on a carbon (``amorphous'') radiator foil to produce an unpolarized photon beam.  

The target used in the g8 run period was a 40\,cm long liquid hydrogen target, located 20\,cm upstream from the geometric center of CLAS. The toroidal magnetic field ran with a current of 1930\,A, which was 50\% of its nominal maximum value and produced a field of roughly 1\,T in the forward region. The polarity of the magnet was set such that positively charged particles were bent outwards, away from the beam axis. The event trigger required a coincidence between a bremsstrahlung electron in the tagging spectrometer and one or more charged particles in CLAS.

The final state particles were detected in the CLAS detector, which was the center-piece of the experimental Hall B at JLab \cite{mecking_cebaf_2003}. CLAS had a six-fold symmetry about the beamline, and consisted of a series of tracking and timing detector subsystems arranged in six sectors. The sectors were separated by superconducting magnet coils that produced a non-uniform toroidal magnetic field of maximum magnitude 1.8\,T. The placement of the detector subsystems led to a particle acceptance polar angle range of 8$^{\circ}$ to 140$^{\circ}$.%

For runs with photon beams, a start counter consisting of scintillator counters surrounding the target region was used to establish a vertex time for an event. Time-of-flight information was measured by a scintillator array and allowed the determination of particle velocities. The deflection of charged particles through the magnetic field was tracked with three regions of drift chambers which, combined with the velocity information from the time-of-flight, were used to deduce the four momentum and charge of the particle. Full details can be found in Ref.\,\cite{mecking_cebaf_2003}.

% === Event Selection =====================================

\section{\label{sec:events}Event Selection}

The reactions of interest in this paper proceed by the following reaction chains:
\[
\vec{\gamma}p\rightarrow K^{+}\Lambda\rightarrow K^{+}p\pi^{-}
\]
\[
\vec{\gamma}p\rightarrow K^{+}\Sigma^{0}\rightarrow K^{+}\gamma\Lambda\rightarrow K^{+}\gamma p\pi^{-},
\]
where the $\Lambda$ and $\Sigma^0$ were measured via the $\Lambda \rightarrow p \pi^-$ decay with 64\% branching ratio. Both two-track ($p, K^+$) and three-track ($p, \pi^-, K^+$) events were retained for further analysis. A comparison between the observables extracted separately from  two-track and three-track events showed that they were consistent within statistical uncertainty, but the final results were extracted with two-track and three-track events combined to optimize accuracy. 

Particle- and channel-identification were performed on data from each coherent edge position.  The photon energy range covered by the coherent peak was $\sim$250\,MeV, resulting in $\sim$50\,MeV overlaps in the data sets relating to each of the different coherent edge positions (1.3, 1.5, 1.7, 1.9, 2.1\,GeV). A comparison of the photon asymmetries in the overlap regions confirmed that the degree of photon polarization had been reliably determined, and extraction of observables was performed on a combined set of all events passing the channel identification criteria. 
% * <david.ireland@glasgow.ac.uk> 2015-09-08T11:12:07.584Z:
%
%  Illustration of the coherent peak overlaps?
%

\subsection{Initial Event Filter}

Since the g8 run period was intended for the measurement of several different channels, the trigger condition was fairly loose. After calibrations had been performed, further analyses on individual channels required a filtering of events (a ``skim'') to reduce the data set to a more manageable number of event candidates. 

Initial particle identification was based on information from the drift chambers, time-of-flight scintillators and the electromagnetic calorimeter. The magnetic field settings meant that the acceptance within CLAS for the negatively charged pion was lower than for the positively charged kaon and proton. For this reason, events with a kaon and a proton were chosen as the best way of reconstructing the hyperon events, with the pion being determined from the missing mass from the $\vec{\gamma}p\rightarrow pK^+X$ reaction. Candidate events required one $pK^+$ pair, with the optional inclusion of a $\pi^-$ and/or neutral particle. These $K^+ \Lambda$ and $K^+ \Sigma^0$ candidates amounted to about 2\% of the total number of recorded events.

\subsection{Particle Identification}
In order to ``clean up'' the remaining data, several other procedures were carried out: a cut to ensure that the particles originated in the hydrogen target; a cut on the relative timing of the photon (as determined by the tagging spectrometer) and the final state hadrons; a cut on the minimum momentum of detected particles; a correction for energy losses in the target and surrounding material; a ``fiducial'' cut to remove events that are detected in regions of CLAS close to the magnet coils and cuts to reduce the background caused by positive pions that are identified as kaons.

A summary of the cuts, together with the effect on the number of surviving reaction channel candidates is given in Table \ref{tab:Analysis-cuts}. 
\begin{table*}
\caption{\label{tab:Analysis-cuts} Analysis cuts applied and resulting number of events for all coherent
peak settings.}
\begin{ruledtabular}
\begin{tabular}{p{5cm}p{7cm}r}
Applied Cut & 
Details &  
Events \\
\colrule
Initial skim                                                        & 
(1 proton) and (1 $K^+$) and (0 or 1 $\pi^-$) and (0 or 1 $\gamma$) & 
$6.03 \times 10^{7}$  \\
Vertex cut on target region                                         & 
$-40<z<0$ cm                                                        & 
$4.71 \times 10^{7}$  \\
$\gamma p$ and $\gamma K^{+}$ vertex timing                         & 
Momentum dependent criterion                                        & 
$1.94 \times 10^{7}$ \\
Minimum momentum cut                                                & 
$p_{p}$ and $p_{K^{+}}>300\,$MeV/c                                  & 
$1.59 \times 10^{7}$ \\
Fiducial cut                                                        & 
$>4^{\circ}$ in azimuthal angle from the sector edges               & 
$1.41 \times 10^{7}$ \\
Pion mis-identification as kaon                                     & 
Assume $p(\gamma,\pi^{+}p)\pi^{-}$, then missing mass $(\pi^{+}p)>\,0.17\,\mathrm{GeV/c^{2}}$                  & 
$9.36 \times 10^6$ \\
Invariant Mass $p\pi^{-}$                      & 
$1.06\,<\,M(p\pi^{-})\,<\,1.2\,\mathrm{GeV/c^{2}}$ & 
$7.06 \times 10^6$ \\
\end{tabular}
\end{ruledtabular}
\end{table*}

\subsection{Channel Identification}

\begin{figure}
\includegraphics[width=0.98\columnwidth]{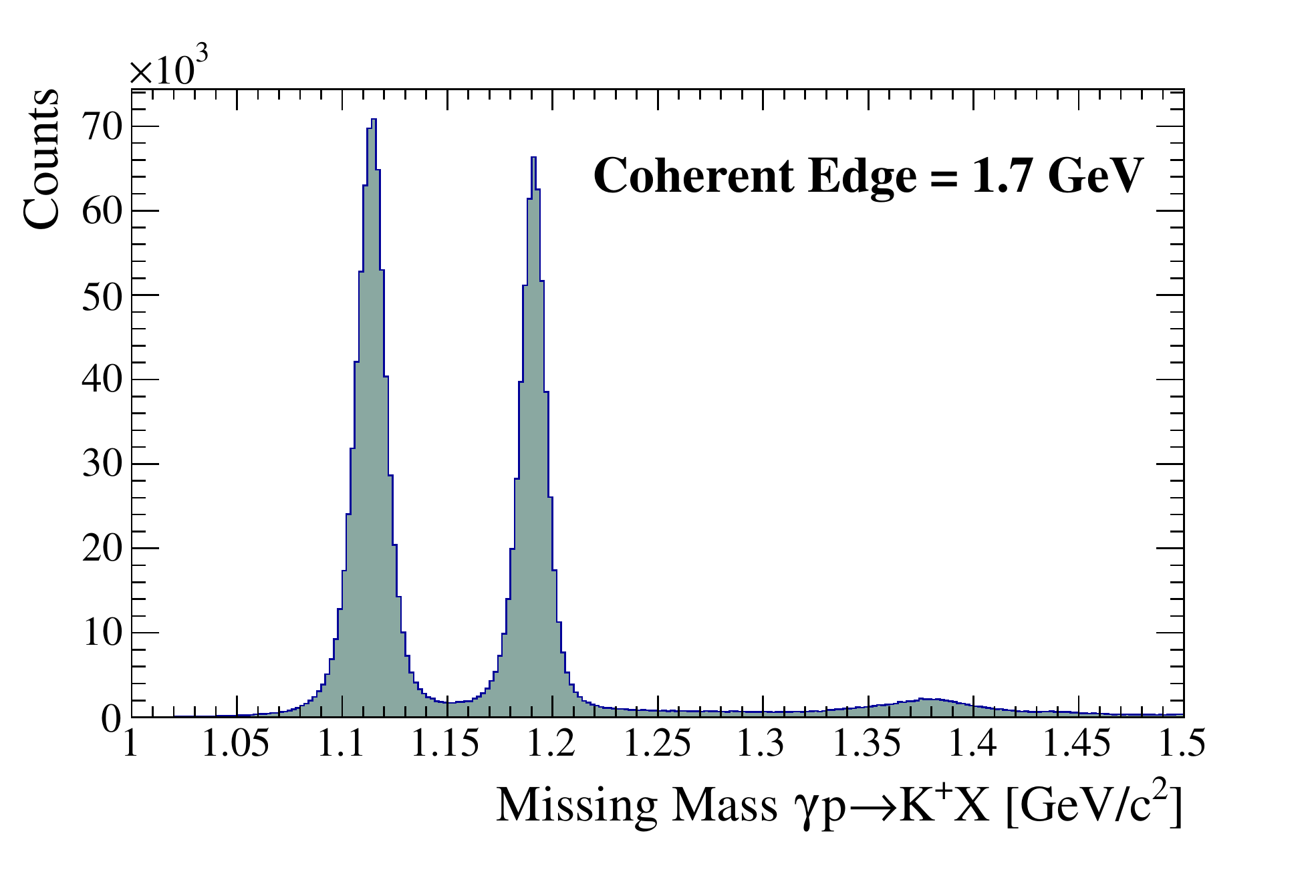}\protect\caption{\label{fig:Missing_Mass}[Color online] Missing mass distribution from the $\vec{\gamma}p\rightarrow K^+X$ reaction. Peaks at 1.115 and 1.193\,GeV/c$^2$ indicate the $\Lambda$ and $\Sigma^0$ events.}
\end{figure}

Figure \ref{fig:Missing_Mass} shows the histogram of missing mass from the $K^+$ for the coherent edge setting of 1.7\,GeV, after the application of the cuts outlined above. Histograms for the other coherent edge settings are almost identical. It is clear from this figure that a very good separation of the $\Lambda$ and $\Sigma^0$ can be made. Note that at a mass of 1.385\,GeV/c$^2$, a bump corresponding to the $\Sigma(1385)$ can be identified. Events with mass within $\pm 2 \sigma$ of the mass of either the $\Lambda$ or the $\Sigma^0$ were retained for further analysis, where $\sigma$ is the standard deviation of the Gaussian part of a Voigtian function (a Lorentzian function convoluted with a Gaussian function). The Lorentzian part has a width parameter $\gamma \ll \sigma$. 

\subsection{\label{sub:photon-beam} Photon Beam Polarization}

In coherent bremsstrahlung \cite{timm_coherent_1969,lohmann_linearly_1994}, the electron beam scatters coherently from a crystal radiator (diamond), resulting in some enhancement over the $\sim1/E_{\gamma}$ photon energy spectrum observed with an amorphous bremsstrahlung radiator. The orientation of the scattering plane is adjusted by setting the azimuthal angle of the crystal lattice in the lab coordinate system. The relative position of the main coherent peak on the photon energy axis is set by adjusting the small angles between the
crystal lattice and the electron beam direction. 

The photons in the coherent peak are linearly polarized and have an angular spread much narrower than that of the unpolarized, incoherent background. By collimating tightly (less than half the characteristic angle), the ratio of polarized to unpolarized photons is increased, and a greater degree of polarization achieved. At typical JLab beam settings (e.g.\,coherent peak $\sim1.3$\,GeV, primary beam $\sim4.5$\,GeV) the degree of linear polarization can be as high as 90\%. In the experiment reported here the range of beam polarization was 50\%---90\%, depending on the photon energy and coherent peak setting.

To measure the degree of polarization in the photon beam, the photon energy spectrum obtained from the tagging spectrometer is fitted with a coherent bremsstrahlung calculation. The parameters of this fit are then used to derive a degree of polarization for the photon beam at intervals of 1\,MeV in photon energy. The fits are performed on every 2 seconds-worth of data, so that a specific degree of polarization can be associated with each event. 

The g8 run period allowed the study of several channels, all of which would be subject to the same systematic uncertainties associated with photon polarization.  As reported in Ref.\,\cite{dugger_beam_2013}, a detailed study of the consistency of the coherent bremsstrahlung calculation was performed, using the reaction $\vec{\gamma} p \rightarrow p\pi^{0}$ \cite{dugger_consistency_2012}. 
In brief, the photon asymmetries in the regions of the overlaps between the coherent peak settings were compared to the published measurements, and the results used to define a small ($<$2\%), energy dependent correction factor. After the application of this correction, we estimate the accuracy of the calculated photon beam polarization to be 3\% for photon energies of 1.9 GeV and below.
At the 2.1\,GeV setting the accuracy was determined to be 6\%. An additional test in Ref.\,\cite{dugger_beam_2013} showed that the systematic uncertainty in the azimuthal angle of the polarization orientation was negligible.

\subsection{\label{sub:background} Background Correction}

It can be seen from Fig.\,\ref{fig:Missing_Mass} that the two hyperons are clearly separated, but that a small residual background has persisted through the various cuts. To estimate the effect of this background, events were divided into 13 bins in $W$ and 4 bins in $\cos\theta^{\star}_K$. A function consisting of a Voigtian function  plus a polynomial background was fitted to the two peaks in each of these bins. There is a small dependence on $W$ and $\cos\theta^{\star}_K$, but the background strength is on average $\lesssim2.5\%$ for the $\Lambda$ and $\lesssim5\%$ for the $\Sigma^{0}$ within the $2\sigma$ cut region.

The background can be accounted for in the extraction of observables, provided that it has no intrinsic asymmetry between events from the parallel and perpendicular settings. We expect this to be the case, since the background is mainly due to uncorrelated pions that just happen to have satisfied the timing cuts. Events falling outside the peak regions in Fig.\,\ref{fig:Missing_Mass} (and associated figures for other coherent edge settings) were examined. Photon beam asymmetries extracted with these events (see Section \ref{sec:observables}) were consistent with zero, and so it was safe to take the fitted background fraction as a simple dilution factor.

% === Extraction of Observables ===========================

\section{\label{sec:observables}Extraction of Observables}

The differential cross section for a pseudo-scalar meson photoproduction experiment can be expressed in terms of sixteen polarization observables, and the degrees of polarization of the beam and target \cite{dey_polarization_2011}. In the case where the photon beam is linearly polarized and the polarization of the recoiling hyperon can be determined via a weak decay asymmetry this reduces to 
\begin{equation}
\begin{array}{rcl}
\dfrac{d\sigma}{d\Omega} & = & \left( \dfrac{d\sigma}{d\Omega}\right)_{0}\left\{ 1-P^{\gamma}\Sigma\cos2\phi\right.\\
 &  & +\alpha\cos\theta_{x}P^{\gamma}O_{x}\sin2\phi\\
 &  & +\alpha\cos\theta_{y}P-\alpha\cos\theta_{y}P^{\gamma}T\cos2\phi\\
 &  & \left.+\alpha\cos\theta_{z}P^{\gamma}O_{z}\sin2\phi\right\}.
\end{array}\label{eq:beam-recoil-1}
\end{equation}
In this expression, $\left( \frac{d\sigma}{d\Omega}\right)_{0}$ represents the unpolarized cross section, $P^{\gamma}$ is the degree of linear photon polarization, $\phi$ is the azimuthal angle between the reaction plane and the photon polarization direction (see Fig.\,\ref{fig:axes}) and  $\Sigma,P,T,O_{x},O_{z}$ are the polarization observables. The direction cosines $\cos\theta_{x,y,z}$ refer to the direction of the decay proton in the hyperon rest frame, and
$\alpha$ is the weak decay asymmetry. The dependence on the kinematic variables $\xi\equiv\left\{ \phi, \cos\theta_x,\cos\theta_y,\cos\theta_z\right\} $ is what allows us to extract the observables.

Note that, since the detection of the proton from the recoiling hyperon is used as a means to identify the channel of interest, measurements will be sensitive to the values of all the observables appearing in Eq.\,(\ref{eq:beam-recoil-1}). It is not possible to ignore any one of the observables by integrating over the decay proton angle; the detection of the proton will automatically bias distributions. It is therefore imperative to extract consistently all the observables to which the experiment is sensitive.

The net result of the preceding channel identification analysis was a selection of events, each of which had a unique set of kinematic variables $\left\{ W, \cos\theta^{\star}_K, \varphi, \cos\theta_x, \cos\theta_y, \cos\theta_z \right\}$, as well as a flag indicating which of the two  settings (parallel or perpendicular) the event came from. The events were sorted into bins of $W$ and $\cos\theta^{\star}_K$, where the binning was defined so that $\gtrsim 1000$ events fell into each bin. 

For each $\left\{ W, \cos\theta^{\star}_K \right\}$ bin, the observables $\left\{ \Sigma,T,O_{x},O_{z}\right\} $  were extracted using an event-by-event asymmetry Maximum Likelihood method.  For each event $e_i$, a likelihood is obtained 
\[
{\cal L}_{i}\left(e_{i}\right) = 
\frac{1}{2}\left(1+\hat{a}_i\right),
\]
where the main ingredient is an estimator of asymmetry:
\begin{equation}
\hat{a}_i=\frac{f_i \Delta L + (1-\beta) P^{\gamma} g_i }
{f_i + (1-\beta) P^{\gamma} g_i \Delta L}.
\label{eq:estimator-main}
\end{equation}
The quantities $P^{\gamma}$, $\Delta L$ and $\beta$ are: degree of photon polarization, asymmetry in the luminosity for each setting (defined as $\left(L_{\perp}-L_{\parallel}\right) / \left( L_{\perp}+L_{\parallel} \right)$) and background fraction, respectively. In the above expression, $f$ and $g$ are derived from the cross section:
\[
\begin{array}{ccl}
f_i  & = & 1+\alpha \cos\theta_{y,i} P\\
g_i  & = & \left(\Sigma+\alpha \cos\theta_{y,i} T\right)\cos2\varphi_i \\
     &   & +\alpha\left(\cos\theta_{x,i} O_{x} + \cos\theta_{z,i} O_{z}\right)\sin2\varphi_i. \\
\end{array}
\]
The details of this derivation and method are left to the appendix.

% === Systematic Uncertainties ============================

\section{\label{sec:systematics}Systematic Uncertainties}

\subsection{Nuisance Parameters}
The quantities $P^{\gamma}$, $\Delta L$ and $\beta$ appearing in Eq.\,(\ref{eq:estimator-main}) represent so-called nuisance parameters, since their values are not intrinsically interesting but do affect the values of extracted observables, and they have to be independently estimated. They therefore represent sources of systematic uncertainty.

As mentioned in Subsection \ref{sub:photon-beam}, the degree of photon linear polarization had an associated systematic uncertainty of 3\% for photon energies up to 1.9\,GeV, whilst data above that energy had a 6\% uncertainty. To estimate the effect of this on the extracted values of observables in $K\Lambda$ and $K\Sigma^{0}$, the extraction procedure was run with values of photon polarization adjusted accordingly. 

The effect of the variation in photon polarization has a noticeable but complicated effect on the extracted values of the observables, due to the  correlations among them. However, the percentage change in photon polarization is roughly equal to the percentage change in the values of the observables, and for the majority of points this systematic uncertainty is less than the statistical uncertainty.

The luminosity asymmetry $\Delta L$ is only dependent on photon energy, and so the procedure to estimate these values was to split the data up into bins in $W$, and perform Maximum Likelihood fits with $\Delta L$ as a free parameter. This was done for events identified as $K\Lambda$ final states and also for events identified as $K\Sigma$ final states. With these two independent means of determining $\Delta L$, the values differed by less than 0.01, and so the uncertainty associated with values of $\Delta L$ was deemed insignificant compared with the statistical accuracy.

As mentioned in Section \ref{sub:background}, the background contribution to the measured events was seen to be $\lesssim 5\%$. The uncertainty on this fitted value was in turn only a few percent, 
so a systematic uncertainty associated with the estimate of the background fraction was ignored.

\subsection{Uncertainties in the Extraction Method}
As mentioned in the appendix, the observables reported here are asymmetries, whose support exists only within the bounds $\left[-1,+1\right]$. To check how imposing this constraint affects the extracted results, we first performed an unconstrained fit (Maximum Likelihood) to check whether there may be systematic uncertainties associated with the evaluation of the nuisance parameters. A constrained fit (maximum posterior probability), which includes the constraint, was then carried out to yield the final numbers. There is no significant difference in the two results from the two methods across the entire kinematic region.

A fraction of the measured events contained final states with three measured particles, which we will refer to as three-track events. A comparison between observables extracted from three-track events ($\pi^-$ detected) and from two-track events ($\pi^-$ reconstructed from missing mass) was carried out. This was done to check both internal consistency, and the calculation of the effective weak decay constant in the case of the $K\Sigma^{0}$ channel \cite{bradford_first_2007}. Both reactions studied here are identified from the detection of a kaon and a proton. In the case of the $K\Lambda$ reaction, this is enough to over-determine the kinematics, whereas the additional photon from the decay of the $\Sigma^0$ means that
there is not a sufficient number of measured kinematic variables to determine
the rest frame of the $\Lambda$, in the decay chain $\Sigma^0\rightarrow\Lambda\gamma;\Lambda\rightarrow\pi^{-}p$. A detailed calculation of how
to extract the $\Sigma^{0}$ polarization components for two-track events
is given in the appendix of \cite{bradford_first_2007}. The values of observables extracted from two- and three-track events in this analysis were all consistent with each other, within the statistical uncertainties.

% === Results =============================================

\section{\label{sec:results}Results}

\begin{figure}
\includegraphics[width=0.58\columnwidth]{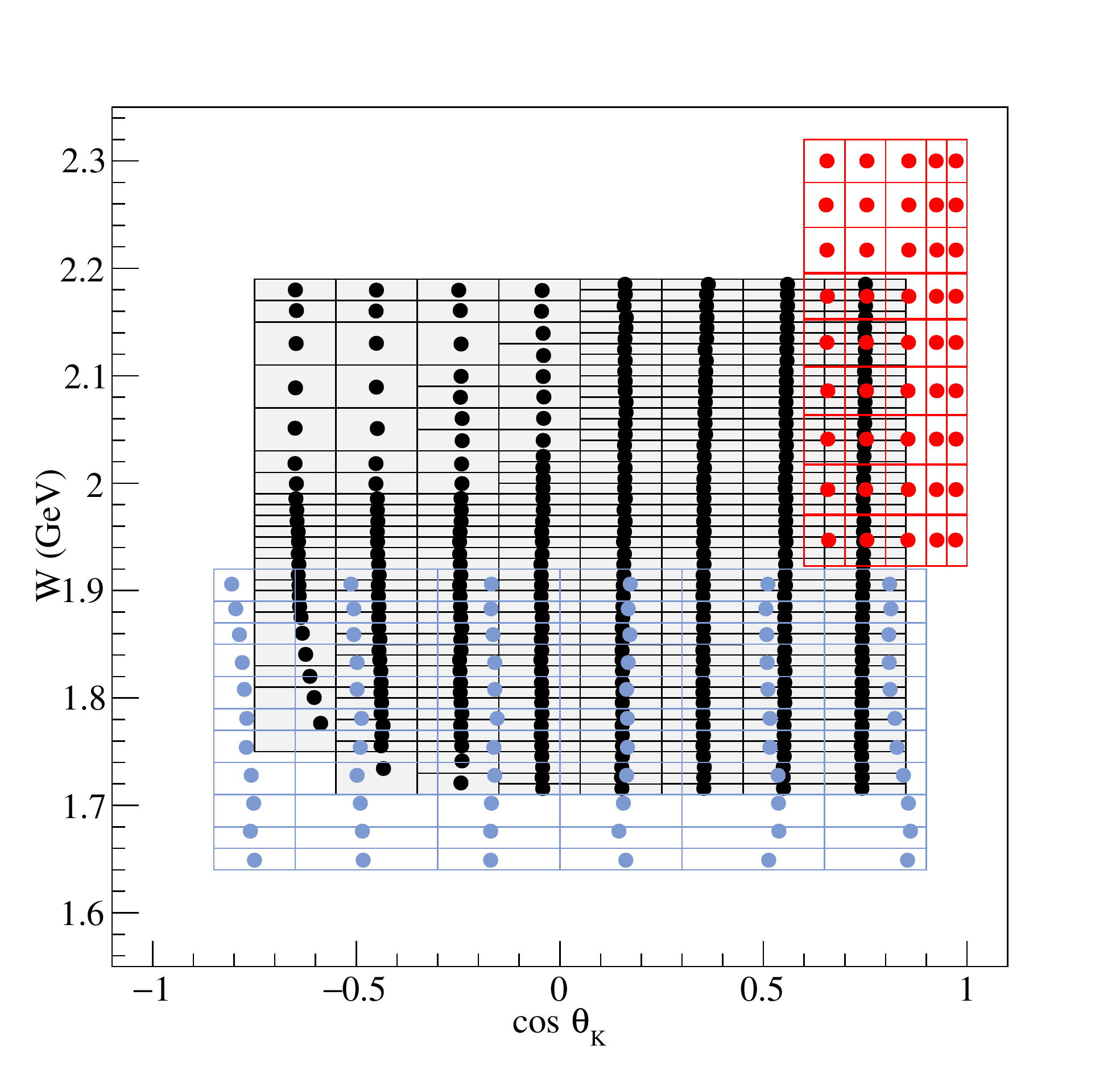}\protect\caption{\label{fig:Comparison-of-kinematic-KL}[Color online] Comparison of kinematic coverage in $W$ vs.\,$\cos\theta^{\star}_K$
for $\vec{\gamma} p\rightarrow K^+\vec{\Lambda}$. Black circles - this (CLAS) measurement; red circles
- LEPS; blue circles - GRAAL. The boxes represent
the limits of the bins in $\left\{ W,\cos\theta^{\star}_K\right\} $.}
\end{figure}

\begin{figure}
\includegraphics[width=0.58\columnwidth]{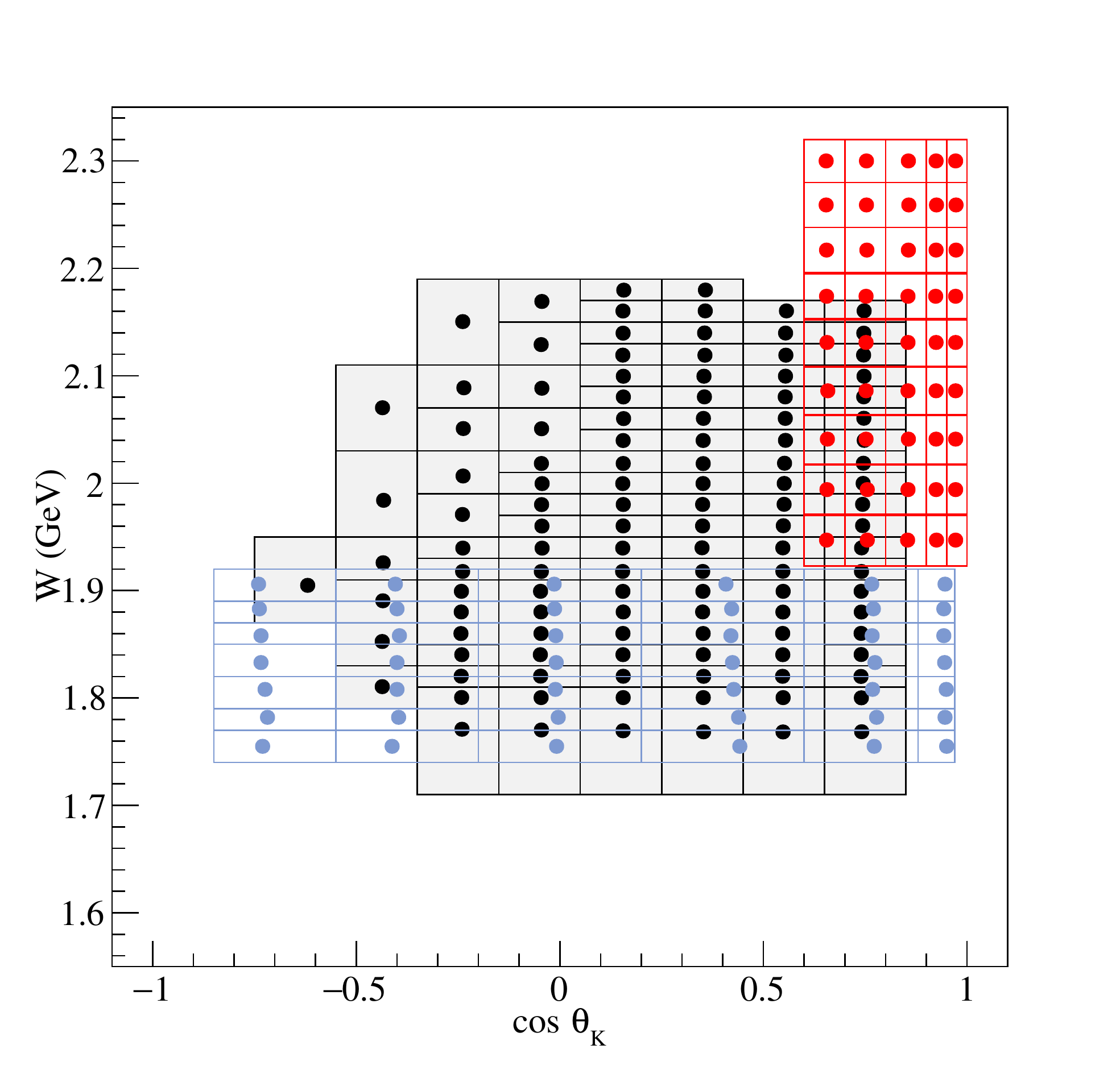}\protect\caption{\label{fig:Comparison-of-kinematic-KS}[Color online] Comparison of kinematic coverage in $W$ vs.\,$\cos\theta^{\star}_K$
for $\vec{\gamma} p\rightarrow K^+\vec{\Sigma}^0$. Black circles - this (CLAS) measurement; red circles
- LEPS; blue circles - GRAAL. The boxes represent
the limits of the bins in $\left\{ W,\cos\theta^{\star}_K\right\} $.}
\end{figure}

The results presented here represent a substantial increase in world data on observables from measurements with linearly polarized photons for the two channels. Figures \ref{fig:Comparison-of-kinematic-KL} and \ref{fig:Comparison-of-kinematic-KS} show the regions in $\left\{ W, \cos\theta^{\star}_K \right\}$ space spanned by the present results, compared to previous ones \cite{lleres_polarization_2007,lleres_measurement_2009,zegers_beam-polarization_2003}. For the CLAS data, the symbols represent the mean value of the bin, weighted by the number of measured events. The symbols for the previous data represent the values reported in the literature \cite{lleres_measurement_2009,lleres_polarization_2007,zegers_beam-polarization_2003}. 
In addition to this, the statistical accuracy of the present data is a significant improvement over the published data in the regions of overlap. A summary of the measurements on the two channels that have been completed so far is given in Table~\ref{tab:Measurements-performed-by}.

\begin{table*} 
\caption{\label{tab:Measurements-performed-by}Measurements performed in the
different experiments.}
\begin{ruledtabular}
\begin{tabular}{llccccccccc}
\textrm{Experiment}&
\textrm{Ref(s)}&
\textrm{Final State}&
\textrm{$W$ range (GeV)}&
$\Sigma$ & $P$ & $C_{x}$ & $C_{z}$ & $T$ & $O_{x}$ & $O_{z}$\\
\colrule
 & & & & & & & & & &\\
CLAS g11 & \cite{mccracken_differential_2010}
		 & $K\Lambda$ & 1.62--2.84 &  & $\star$ &  &  &  &  & \\
         & \cite{dey_differential_2010}
         & $K\Sigma^{0}$ & 1.69--2.84 &  & $\star$ &  &  &  &  & \\
\colrule
 & & & & & & & & & &\\
CLAS g1c & \cite{mcnabb_hyperon_2004,bradford_first_2007}
         & $K\Lambda$ & 1.68--2.74 &  & $\star$ & $\star$ & $\star$ &  &  & \\
         & \cite{mcnabb_hyperon_2004,bradford_first_2007}         
         & $K\Sigma^{0}$ & 1.79--2.74 &  & $\star$ & $\star$ & $\star$ &  &  & \\ 
\colrule         
 & & & & & & & & & &\\
LEPS     & \cite{zegers_beam-polarization_2003}
         & $K\Lambda$ & 1.94--2.30 & $\star$ &  &  &  &  &  & \\
         & \cite{zegers_beam-polarization_2003}
         & $K\Sigma^{0}$ & 1.94--2.30 & $\star$ &  &  &  &  &  & \\
\colrule
 & & & & & & & & & &\\
GRAAL    & \cite{lleres_polarization_2007,lleres_measurement_2009}
         & $K\Lambda$ & 1.64--1.92 & $\star$ & $\star$ &  &  & $\star$ & $\star$ & $\star$\\
         & \cite{lleres_polarization_2007}
         & $K\Sigma^{0}$ & 1.74--1.92 & $\star$ & $\star$ &  &  &  &  & \\
\colrule
 & & & & & & & & & &\\
CLAS g8  &
         & $K\Lambda$ & 1.71--2.19 & $\star$ & $\star$ &  &  & $\star$ & $\star$ & $\star$\\
         &         
         & $K\Sigma^{0}$ & 1.75--2.19 & $\star$ & $\star$ &  &  & $\star$ & $\star$ & $\star$\\
\end{tabular}
\end{ruledtabular}
\end{table*}

\begin{figure*}
\includegraphics[width=0.98\textwidth]{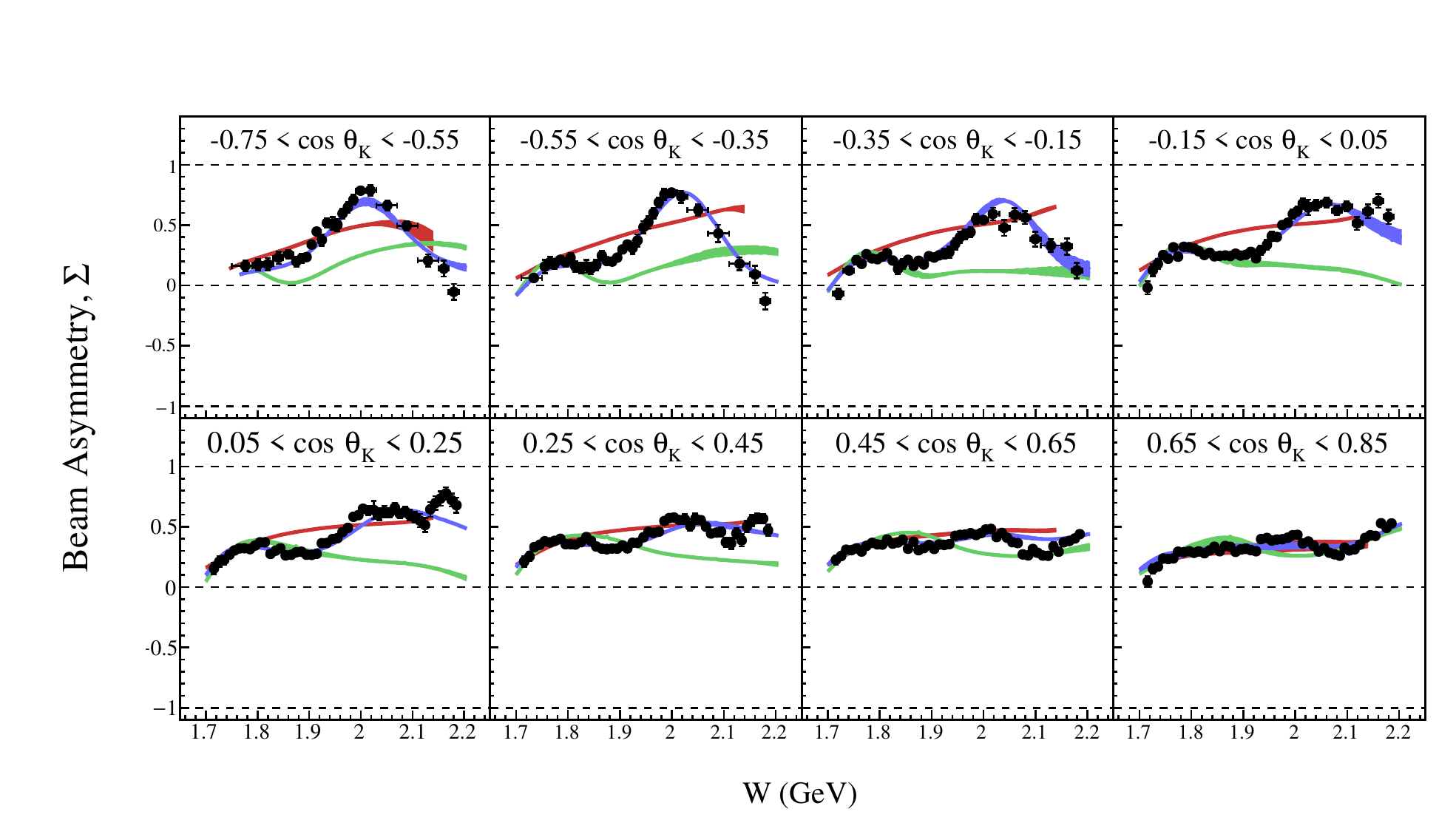}
\protect\caption{\label{fig:Beam-asymmetry-KL}[Color online] The energy dependence of the beam asymmetry, $\Sigma$, for the reaction $\vec{\gamma} p\rightarrow K\vec{\Lambda}$. Red curves - ANL-Osaka predictions from coupled-channels calculations \cite{kamano_nucleon_2013}; Green curves - predictions from the 2014 solution of the Bonn-Gatchina partial wave analysis \cite{gutz_high_2014}; Blue curves - Bonn-Gatchina calculations after a re-fit including the present data, which include additional $N^{\star}(\frac{3}{2}^+)$ and $N^{\star}(\frac{5}{2}^+)$ resonances \cite{sarantsev_private_2015}}.
\end{figure*}

\begin{figure*}
\includegraphics[width=0.98\textwidth]{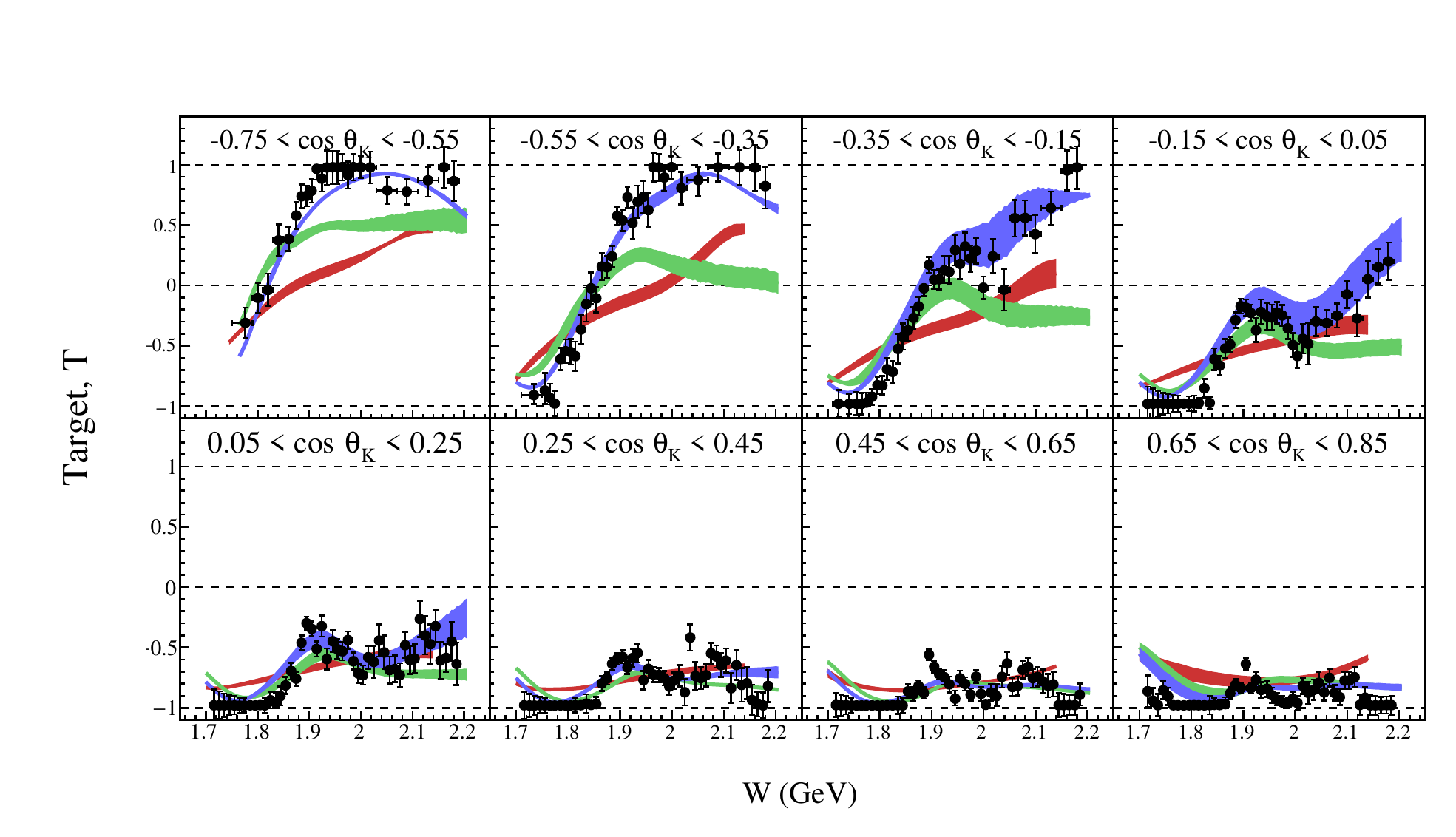}\protect\caption{\label{fig:Target-asymmetry-KL}[Color online] The energy dependence of the target asymmetry, $T$, for the reaction $\vec{\gamma} p\rightarrow K\vec{\Lambda}.$ The curves have the same definition as in Fig.\,\ref{fig:Beam-asymmetry-KL}.}
\end{figure*}

\begin{figure*}
\includegraphics[width=0.98\textwidth]{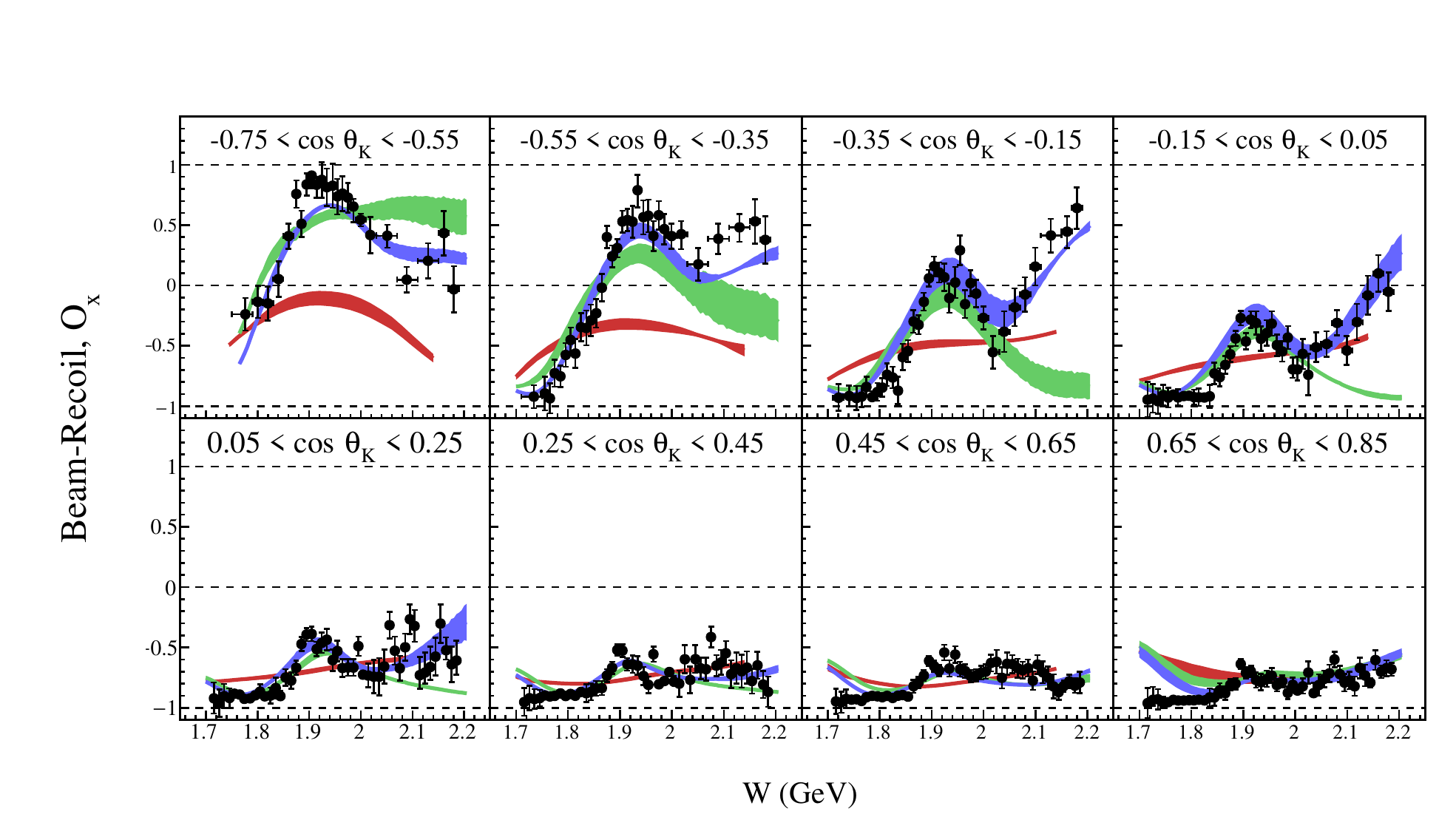}\protect\caption{\label{fig:Ox-beam-recoil-KL}[Color online] The energy dependence of the beam-recoil double asymmetry, $O_x$, for the reaction $\vec{\gamma} p\rightarrow K\vec{\Lambda.}$ The curves have the same definition as in Fig.\,\ref{fig:Beam-asymmetry-KL}.}
\end{figure*}

\begin{figure*}
\includegraphics[width=0.98\textwidth]{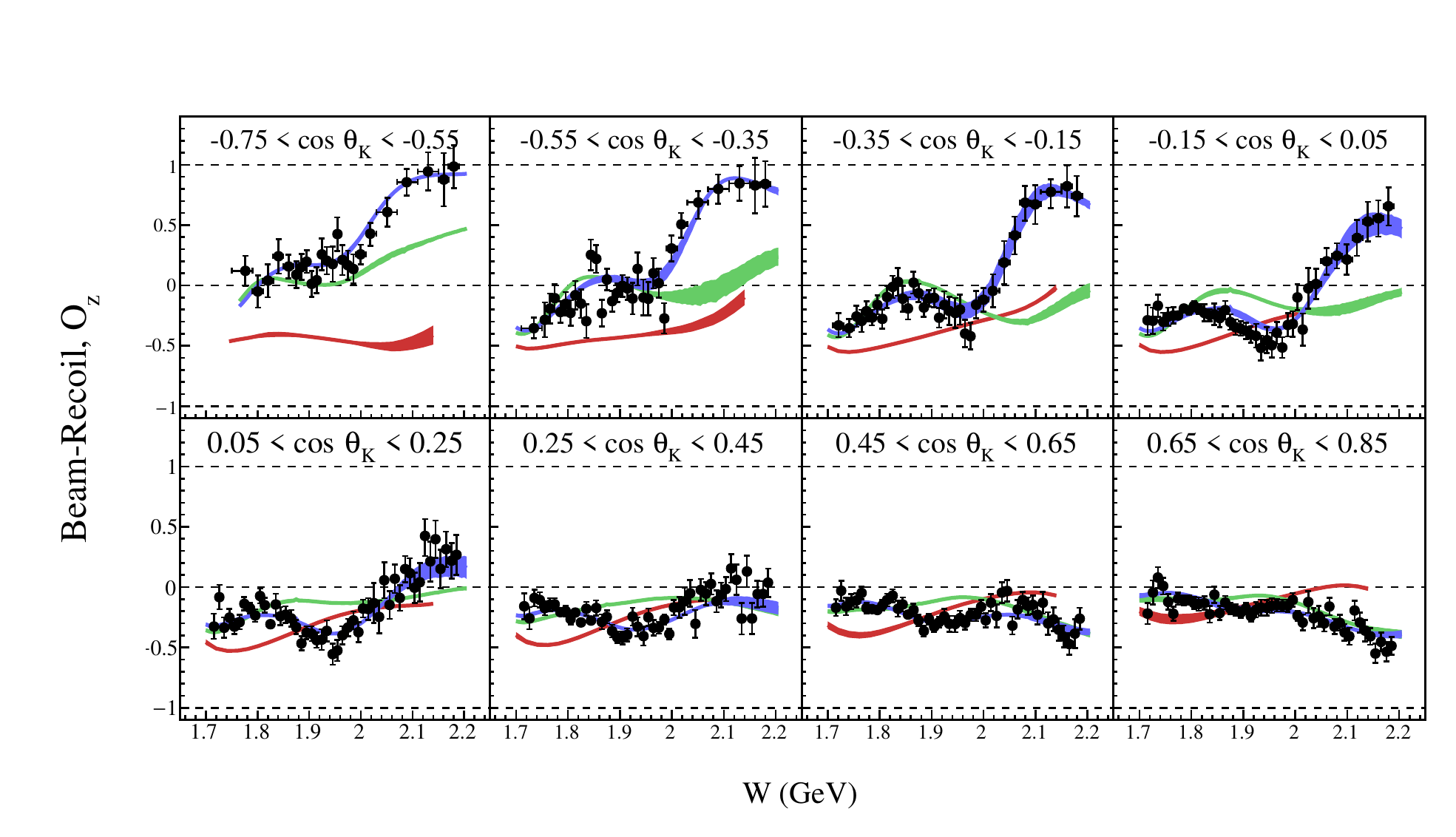}\protect\caption{\label{fig:Oz-beam-recoil-KL}[Color online] The energy dependence of the beam-recoil double asymmetry, $O_z$, for the reaction $\vec{\gamma} p\rightarrow K\vec{\Lambda}.$ The curves have the same definition as in Fig.\,\ref{fig:Beam-asymmetry-KL}.}
\end{figure*}

\begin{figure*}
\includegraphics[width=0.98\textwidth]{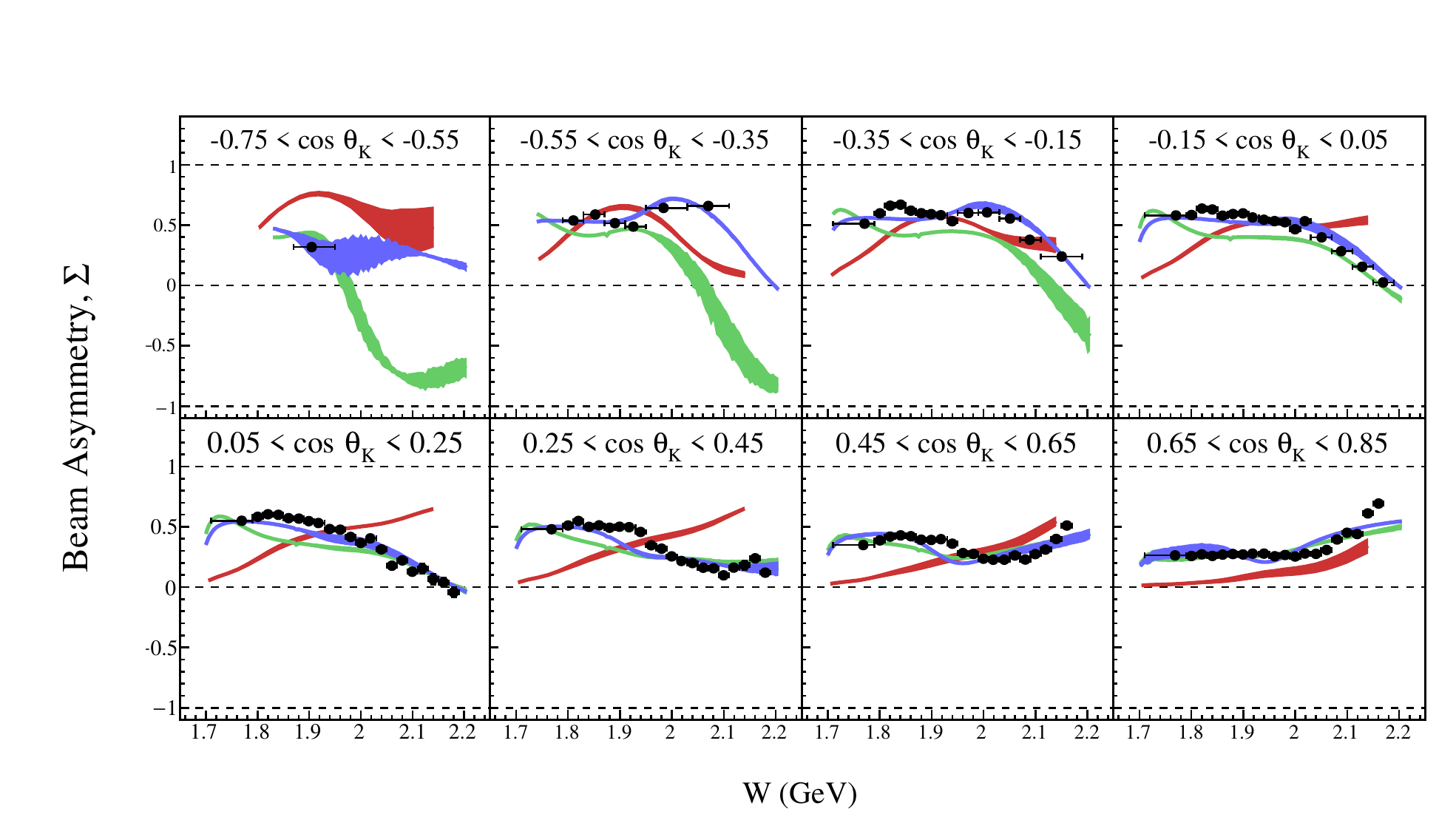}\protect\caption{\label{fig:Beam-asymmetry-KS}[Color online] The energy dependence of the beam asymmetry, $\Sigma$, for the reaction $\vec{\gamma} p\rightarrow K\vec{\Sigma}^0.$ The curves have the same definition as in Fig.\,\ref{fig:Beam-asymmetry-KL}.}
\end{figure*}

\begin{figure*}
\includegraphics[width=0.98\textwidth]{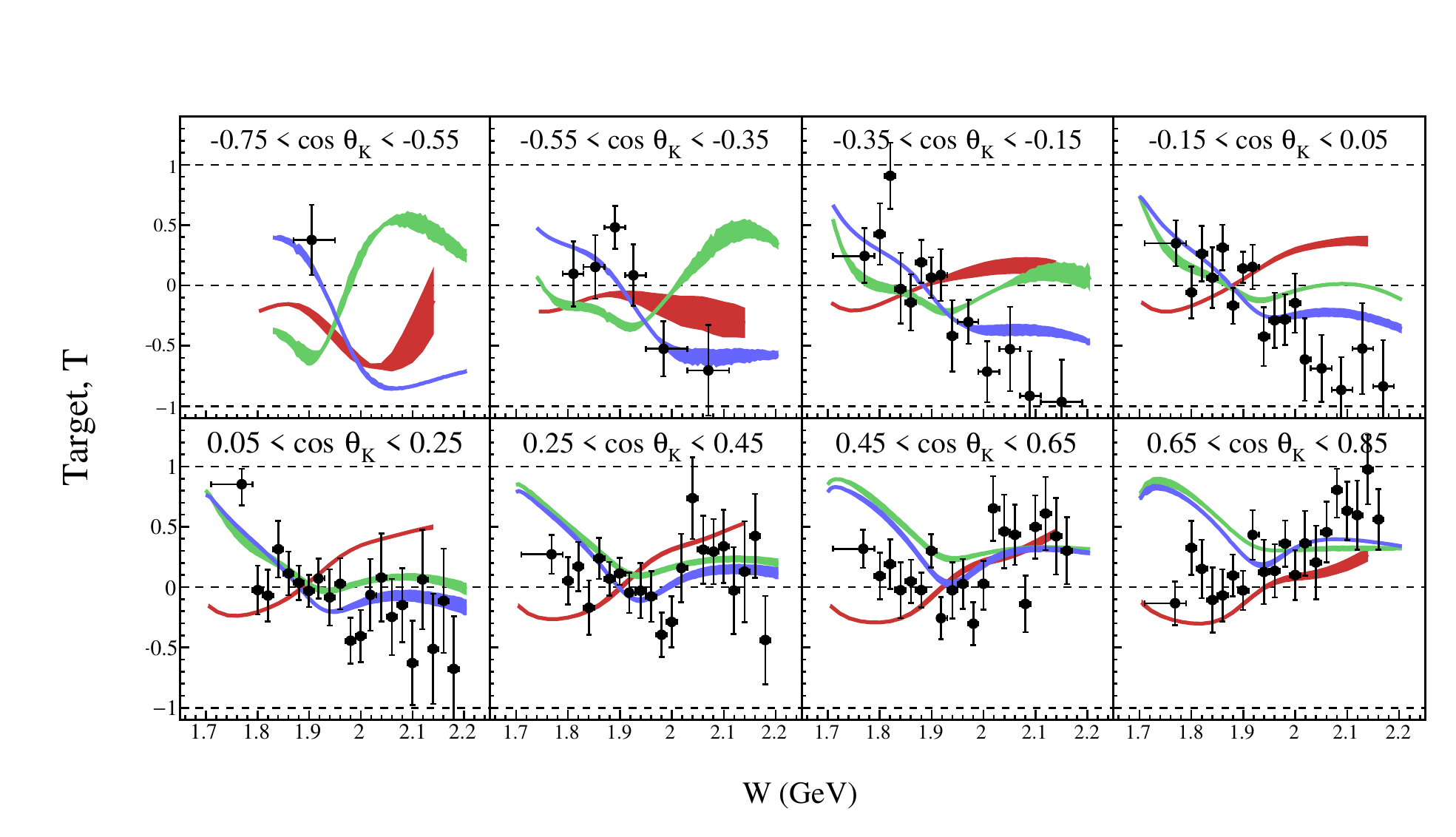}\protect\caption{\label{fig:Target-asymmetry-KS}[Color online] The energy dependence of the target asymmetry, $T$, for the reaction $\vec{\gamma} p\rightarrow K\vec{\Sigma}^0.$ The curves have the same definition as in Fig.\,\ref{fig:Beam-asymmetry-KL}.}
\end{figure*}

\begin{figure*}
\includegraphics[width=0.98\textwidth]{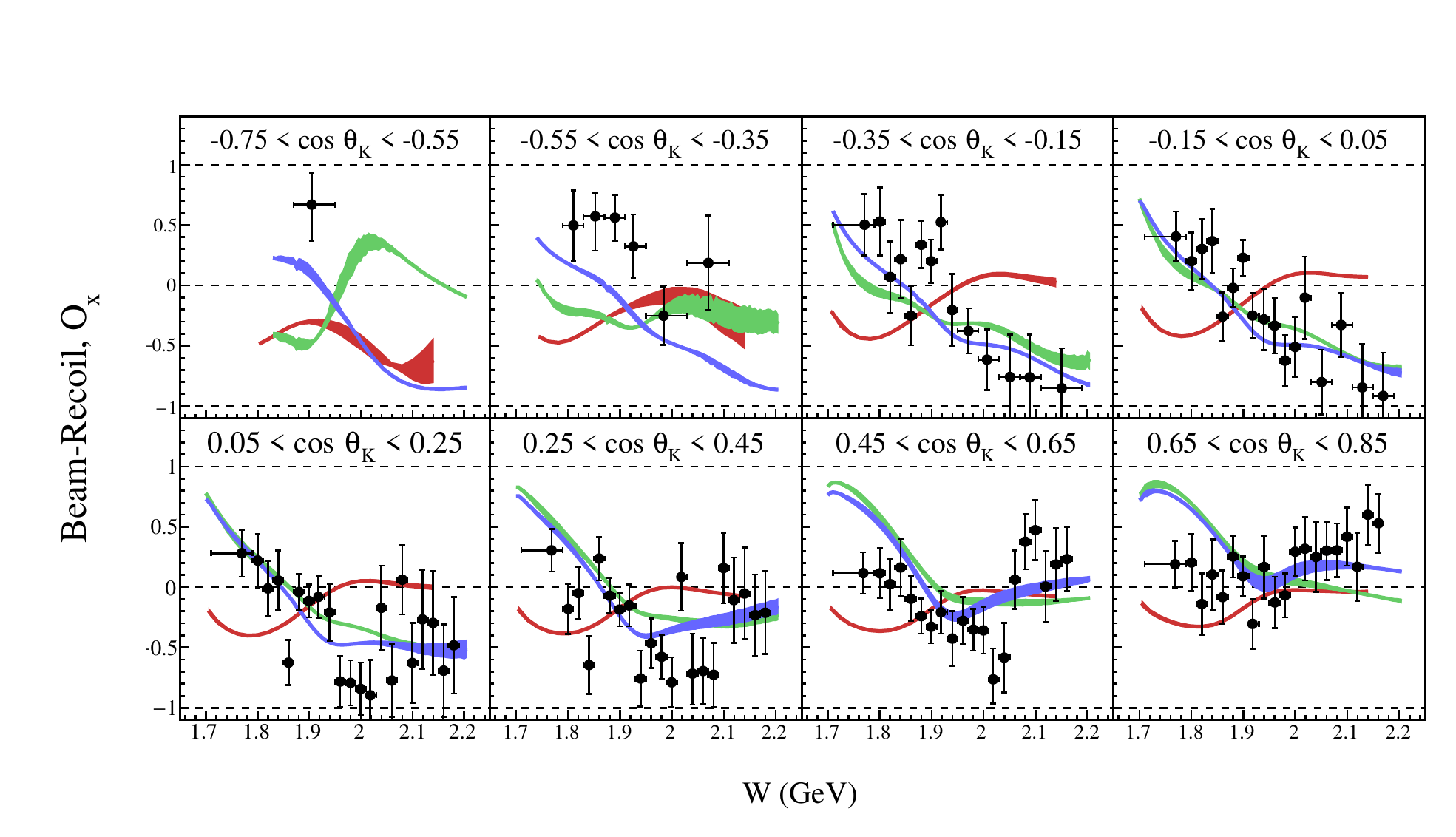}\protect\caption{\label{fig:Ox-beam-recoil-KS}[Color online] The energy dependence of the beam-recoil double asymmetry, $O_x$, for the reaction $\vec{\gamma} p\rightarrow K\vec{\Sigma}^0.$ The curves have the same definition as in Fig.\,\ref{fig:Beam-asymmetry-KL}.}
\end{figure*}

\begin{figure*}
\includegraphics[width=0.98\textwidth]{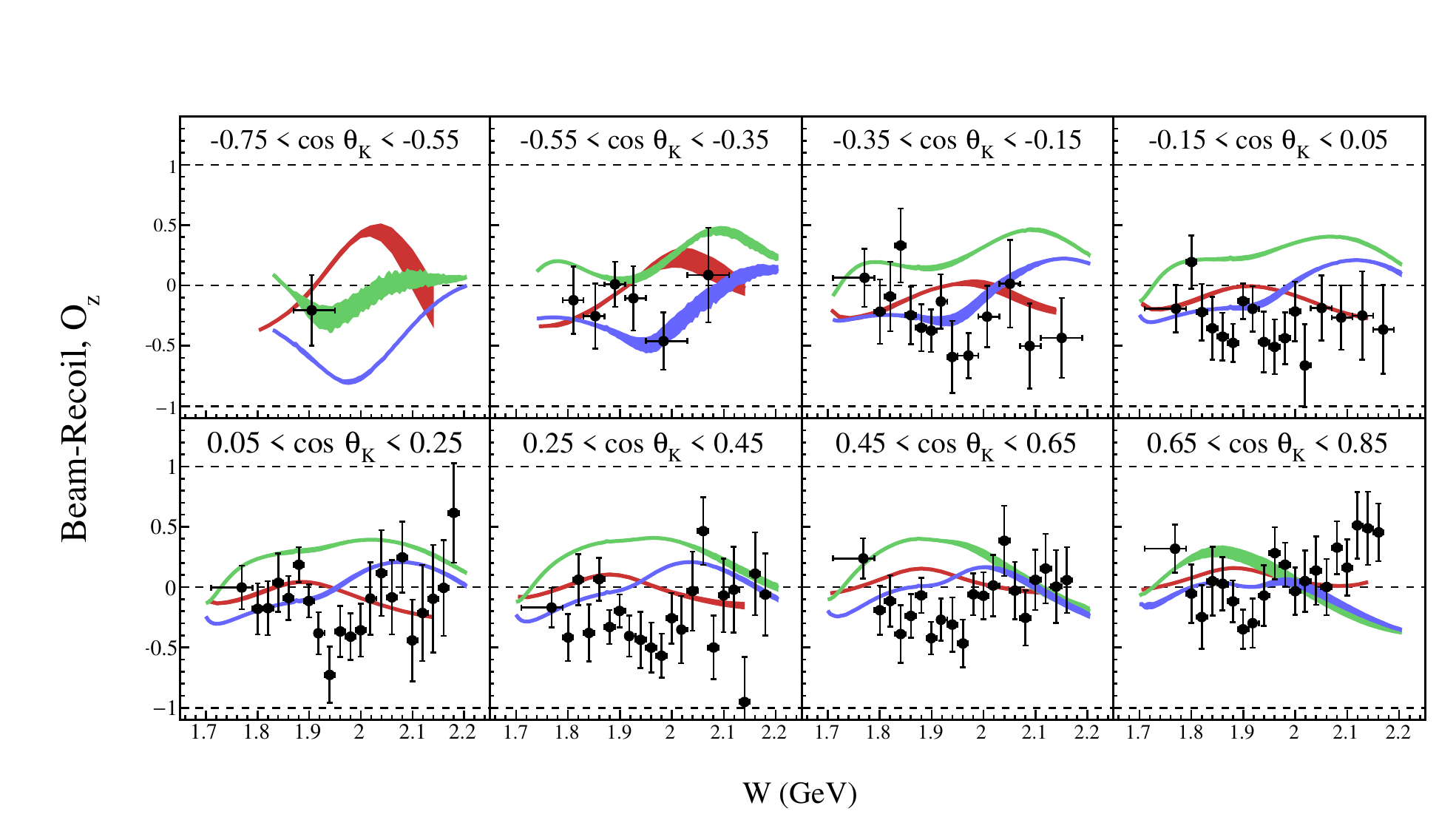}\protect\caption{\label{fig:Oz-beam-recoil-KS}[Color online] The energy dependence of the beam-recoil double asymmetry, $O_z$, for the reaction $\vec{\gamma} p\rightarrow K\vec{\Sigma}^0.$ The curves have the same definition as in Fig.\,\ref{fig:Beam-asymmetry-KL}.}
\end{figure*}

The results for the observables $\left\{ \Sigma,T,O_{x},O_{z}\right\} $ for the $\vec{\gamma}  p \rightarrow K^+ \vec{\Lambda}$ reaction are displayed in Figs.\,\ref{fig:Beam-asymmetry-KL}-\ref{fig:Oz-beam-recoil-KL}, while the same observables for the $\vec{\gamma}  p \rightarrow K^+ \vec{\Sigma}^0$ reaction are shown in Figs.\,\ref{fig:Beam-asymmetry-KS}-\ref{fig:Oz-beam-recoil-KS} \footnote{Numerical results are publicly available from the CLAS Physics Database \texttt{ http://clasweb.jlab.org/physicsdb}, or on request from the corresponding author \texttt{David.Ireland@glasgow.ac.uk}.}. Where visible, horizontal bars on the data indicate the angular limits of the bins, corresponding to those illustrated in Figs.\,\ref{fig:Comparison-of-kinematic-KL} and \ref{fig:Comparison-of-kinematic-KS}. 

Also shown in the figures are three calculations. The red curves show predictions from the ANL-Osaka group \cite{kamano_nucleon_2013}, which are dynamical coupled-channels calculations incorporating known resonances with masses below 2 GeV/c$^2$, which have total widths less than 400 MeV/c$^2$ and whose pole positions and residues could be extracted. The green curves represent predictions from the 2014 solution of the Bonn-Gatchina partial wave analysis (BG2014-02, \cite{gutz_high_2014}), whilst the blue curves are the result of a re-fit solution of the Bonn-Gatchina partial wave analysis \cite{sarantsev_private_2015} of data from all channels, including the new data reported here.

For a comparison of the calculations with the data, calculations from each of the groups were supplied on a fine grid in $W$ and $\cos\theta^{\star}_K$. Each CLAS data point represents a weighted average of the observable in a finite bin of $W$ and $\cos\theta^{\star}_K$. A weighted average of the calculations that took into account the distribution of measured events within the bin was evaluated. The bands observed in the plots represents the standard deviation of calculations within the kaon angular range labelled in the sub-plots.

It is clear from the plots that there is a great deal of structure in the $W-$ and $\cos\theta^{\star}_K -$ dependence of each of the observables. For the two calculations that represent predictions (ANL-Osaka and Bonn-Gatchina-2014), the fits generally appear to match the data reasonably well at forward angles over most of the energy range, and for $W<1.8$\,GeV at backward angles over most of the angular range. These ranges in $\left\{ W, \cos\theta^{\star}_K \right\}$ space are where the data sets from LEPS and GRAAL were used in the previous theoretical fits. Away from the regions that overlap with the previous data, however, these predictions do not do well in matching the data. The re-fit of the Bonn-Gatchina solution does indicate a good agreement over the whole kinematic region for the $K-\Lambda$ channel, and fair agreement for the $K-\Sigma$ channel. 

For the Bonn-Gatchina re-fit, the resonance set in the BG2014-02 solution was used, and data from all two-body final states were fitted. In doing this, the couplings to three-body final states were held fixed, while all other parameters were allowed to vary. This resulted in a reasonable description of all data, and was used as a baseline for further studies. The fact that this fit was able to reproduce the present data, and all previous data, satisfactorily can be attributed to the fact that very small differences in some parameters, such as phases, can give rise to large differences in some observable quantities in one channel, without greatly affecting other channels.

A comprehensive program of including one or two additional resonances in the mass region 2.1-2.3\,GeV\/c$^2$ was undertaken. Several hundred new fits were performed, each one of which involved the introduction of a combination of states with a variety of spins and parities. Of these, an overall best fit was found with the addition of two new resonances: $N^{\star}(\frac{3}{2}^+)$ and $N^{\star}(\frac{5}{2}^+)$. However the improvement in fit was not significant enough to determine their masses, or indeed to claim strong evidence for their existence. There were many combinations that showed small improvements in goodness-of-fit, and so the conclusion is that the new data are suggestive of additional resonances, but further data will be required to establish their identities. 

The re-fit curves shown in the plots are calculations that include the additional $N^{\star}(\frac{3}{2}^+)$ and $N^{\star}(\frac{5}{2}^+)$ states. However, the difference between these distributions and those corresponding to the fit with no new resonances is not possible to discern on the graphs; the improvement in the fit is small and is also ``diluted'' over several channels and many observables. 

The ``predictive power'' of the BG2014-02 solution appears to have been poor in the regions where there has previously been no data. However, this approach to fitting data from many channels is less about developing a predictive model, and more about being able to extract more information from data when more data are available. It is a further indication that polarization observables of sufficient accuracy will indeed be required to extract the full physics information from these channels \cite{chiang_completeness_1997,ireland_information_2010}. 

As a check of consistency with previous measurements, we can make use
of one of several identities that connect the polarization observables
for pseudoscalar meson photoproduction \cite{sandorfi_determining_2011}, known as the ``Fierz identities''. 

Previous CLAS measurements of the $K\Lambda$ and $K\Sigma^{0}$ channels have reported: differential cross sections and recoil polarizations \cite{bradford_first_2007,mccracken_differential_2010,dey_differential_2010}; circular beam-recoil observables $C_{x}$ and $C_{z}$ \cite{bradford_first_2007}.
The measurements were all taken in a similar range of $W$ and $\cos\theta^{\star}_K$ to the work reported here. The relation
\[
O_{x}^{2}+O_{z}^{2}+C_{x}^{2}+C_{z}^{2}+\Sigma^{2}-T^{2}+P^{2}=1
\]
connects all the observables measured in the CLAS experiments (relation labelled S.br in ref.\,\cite{sandorfi_determining_2011}). We can therefore compare
$C_{x}^{2}+C_{z}^{2}$ from  \cite{bradford_first_2007} with the combination $1-O_{x}^{2}-O_{z}^{2}-\Sigma^{2}+T^{2}-P^{2}$ measured here, where the value of $P$ used is an interpolation of results in  \cite{mccracken_differential_2010,dey_differential_2010}.  

\begin{figure*}
\includegraphics[width=0.98\textwidth]{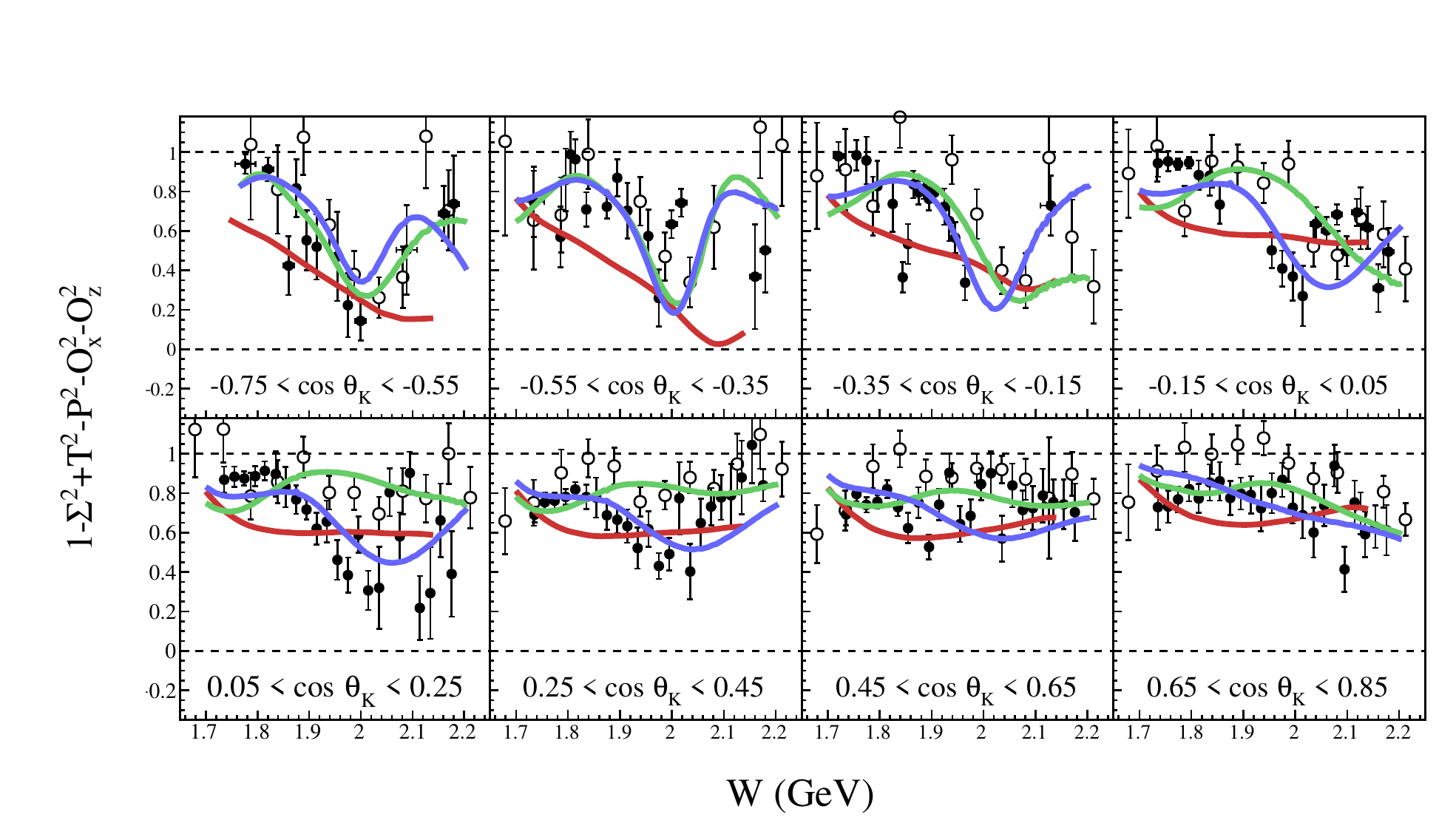}\protect\caption{\label{fig:Fierz-KL} [Color online] Comparison of the combination of present measurements $1-O_{x}^{2}-O_{z}^{2}-\Sigma^{2}+T^{2}-P^{2}$ (black circles) with the combination of previous beam-recoil measurements $C_{x}^{2}+C_{z}^{2}$ (open circles \cite{bradford_first_2007}) to check a Fierz identity. The colored lines represent the values of the combination as evaluated from the three theoretical models described earlier (Fig.\,\ref{fig:Beam-asymmetry-KL}).}
\end{figure*}

The results of the comparison are shown in Fig.\,\ref{fig:Fierz-KL}, together with the values derived from the theoretical models that have been compared to the individual observables. By definition, the combinations $C_{x}^{2}+C_{z}^{2}$  and $1-O_{x}^{2}-O_{z}^{2}-\Sigma^{2}+T^{2}-P^{2}$ from the models are equal. 

Whilst the error bars from the combinations are large, the two data sets are not inconsistent with each other.  Note that in the present work, \emph{all} the $\Sigma,P,T,O_{x}, O_{z}$ observables are extracted at once and have been constrained to the physical region, whereas in the previous work, the $C_x$ and $C_z$ observables were extracted independently and were not constrained to the physical region.

% === Conclusions  ========================================

\section{\label{sec:conclusions}Conclusions}

Measurements of polarization observables for the reactions $\vec{\gamma} p \rightarrow K^+ \Lambda$ and $\vec{\gamma} p \rightarrow K^+ \Sigma^0$ have been performed. The energy range of the results is 1.71\,GeV $<W<$ 2.19\,GeV, with an angular range $-0.75 < \cos\theta^{\star}_K < +0.85$. The observables extracted for both reactions are beam asymmetry $\Sigma$, target asymmetry $T$, and the beam-recoil double polarization observables $O_x$ and $O_z$. This greatly increases the world data set for the observables in the $\vec{\gamma} p \rightarrow K^+ \Lambda$ channel, both in kinematic coverage and in accuracy. The $T$, $O_x$ and $O_z$ data for the $\vec{\gamma} p \rightarrow K^+ \Sigma^0$ channel are new, and the beam asymmetry measurements also increase kinematic coverage and accuracy over previous measurements.

Comparison with phenomenological fits of the Bonn-Gatchina model indicate that this present data set shows some evidence of resonances beyond the 2014 solution, but that it is not strong enough to deduce the quantum numbers or masses of these states or indeed conclusively support their existence. Comparison with the ANL-Osaka calculations indicate that this model may not include sufficient resonance information. Further data, including additional polarization observables and results from other channels, are being analyzed and will be the subject of future CLAS publications. Such data will still be necessary to untangle the full spectrum of $N^{\star}$ resonances.

% === Acknowledgements  ===================================

\begin{acknowledgments}
The authors gratefully acknowledge the work of Jefferson Lab staff in the Accelerator and Physics Divisions.
This work was supported by: the United Kingdom's Science and Technology Facilities Council (STFC) from grant numbers ST/F012225/1, ST/J000175/1 and ST/L005719/1;
the Chilean Comisi\'on Nacional de Investigaci\'on Cient\'ifica y Tecnol\'ogica (CONICYT);
the Italian Istituto Nazionale di Fisica Nucleare;
the French Centre National de la Recherche Scientifique;
the French Commissariat \`{a} l'Energie Atomique;
the U.S.~National Science Foundation;
the National Research Foundation of Korea.
Jefferson Science Associates, LLC, operates the Thomas Jefferson National Accelerator Facility for the the U.S.~Department of Energy under contract DE-AC05-06OR23177.
We also thank Andrei Sarantsev for providing calculations from the re-fit Bonn-Gatchina partial wave analysis.
\end{acknowledgments}

% === Appendix  ===========================================

\appendix

\section{\label{app:subsec}Extraction of Polarization Observables}

A method for estimating the values of observables was developed, which used event-by-event Maximum Likelihood fits to data sorted into bins in $W$ and $\cos\theta^{\star}_K$. While there are numerous examples of event based likelihood fits (either Maximum Likelihood or Extended Maximum Likelihood), this methodology has not been used for asymmetry measurements before, so we outline the procedure in this appendix.

The cross section, as defined in Eq.\,(\ref{eq:beam-recoil-1}), is a function of the hadronic mass $W$ and the center of mass kaon scattering angle $\theta^{\star}_K$. The rest of this appendix assumes that we are discussing one bin in $W$ and $\cos\theta^{\star}_K$. We can re-write the cross section as:
\begin{equation}
\sigma^{s}_{\perp (\parallel)} =
\sigma_0 \left( f - P^{\gamma}_{\perp (\parallel)} 
g_{\perp (\parallel)} \right),
\label{eq:cs}
\end{equation}
where
\begin{equation}
\begin{array}{ccl}
f             & = & 1+\alpha \cos\theta_y P\\
g_{\perp}     & = & -\left(\Sigma+\alpha \cos\theta_y T\right)\cos2\varphi \\
              &   & -\alpha\left(\cos\theta_x O_{x} + \cos\theta_z O_{z}\right)\sin2\varphi \\
g_{\parallel} & = & +\left(\Sigma+\alpha \cos\theta_y T\right)\cos2\varphi \\
              &   & +\alpha\left(\cos\theta_x O_{x} + \cos\theta_z O_{z}\right)\sin2\varphi. \\
\end{array}\label{eq:f-and-g}
\end{equation}
The effect of changing settings is to reverse the sign in front of the sine and cosine terms, so we can write
\[
g_{\parallel} = - g_{\perp} = g.
\]
Also, the superscript $s$ is used to denote the cross section for \emph{signal} events.

Within one $\left\{ W,\cos\theta^{\star}_K\right\} $ bin, there is a distribution in the variables $\xi\equiv\left\{ \phi, \cos\theta_x,\cos\theta_y,\cos\theta_z\right\} $, the form
of which allows us to estimate the polarization observables. Throughout
such a bin, we assume that there is a true asymmetry $a\left(\xi\right)\in\left[-1,1\right]$. In a specified range of $\xi$, the probability of obtaining exactly $n_{\perp}$ and $n_{\parallel}$ counts in the perpendicular and parallel settings respectively, given a specific value of $a$ would be
\begin{equation}
{\cal P}\left(n_{\perp},n_{\parallel}\mid a\right)=\frac{1}{Z}\left(1+a\right)^{n_{\perp}}\left(1-a\right)^{n_{\parallel}},\label{eq:bin-likelihood}
\end{equation}
where $Z$ is a normalizing constant. 

In an event-by-event analysis, the range in $\xi$ is such that it contains just one event. Events can be denoted by 
\[
e_{\perp}     \equiv \left\{ n_{\perp}=1,n_{\parallel}=0 \right\};\quad
e_{\parallel} \equiv \left\{ n_{\perp}=0,n_{\parallel}=1 \right\}.
\]

Equation (\ref{eq:bin-likelihood}) would then become either of:
\begin{equation}
{\cal P}\left( e_{\perp} \mid a \right)=\frac{1}{2}\left(1+a\right);\quad
{\cal P}\left( e_{\parallel} \mid a \right)=\frac{1}{2}\left(1-a\right),
\label{eq:bin-perp}
\end{equation}
depending on the setting.

We now need to construct an estimator $\hat{a}$ for the asymmetry. It will be a function of the variables $\xi$, but will also depend on the observables of interest, ${\cal O}\equiv\left\{ \Sigma,P,T,O_{x},O_{z}\right\} $, and other quantities referred to as ``nuisance parameters'' $\lambda$. These nuisance parameters represent quantities, such as degree of photon polarization, that must be determined independently and give rise to systematic uncertainties.

The measured number of counts in each setting will be related to the detector acceptance, the integrated luminosity and the cross section, so the expected numbers will be:
\[
\overline{n}_{\perp (\parallel)} = \varepsilon_{\perp (\parallel)} L_{\perp (\parallel)} \sigma^{c}_{\perp (\parallel)}.
\]
$\varepsilon_{\perp (\parallel)}$ is the acceptance and $L_{\perp (\parallel)}$ the luminosity. The superscript $c$ in the cross section symbols indicates that the cross section is a combination of both signal $s$ and background $b$:
\[
\sigma^{c}_{\perp (\parallel)} = 
\sigma^{s}_{\perp (\parallel)} + \sigma^{b}, 
\]
where it is assumed that the background contribution does not depend on photon polarization setting (as shown in Section \ref{sub:background}). The expected asymmetry of counts is then:
\begin{equation}
\overline{\Delta n} = \frac{\overline{n}_{\perp}-\overline{n}_{\parallel}}{\overline{n}_{\perp}+\overline{n}_{\parallel}}
= \frac{\varepsilon_{\perp} L_{\perp} \sigma^{c}_{\perp} -\varepsilon_{\parallel} L_{\parallel} \sigma^{c}_{\parallel}}{\varepsilon_{\perp} L_{\perp} \sigma^{c}_{\perp} +
\varepsilon_{\parallel} L_{\parallel} \sigma^{c}_{\parallel}}.\label{eq:count-asym}
\end{equation}
The detector does not measure the photon polarization direction, so the acceptance for a phase-space volume in both settings is the same; it can therefore be divided out.

Taking the asymmetries of cross sections and luminosities:
\[
\Delta \sigma=\frac{\sigma^c_{\perp}-\sigma^c_{\parallel}}{\sigma^c_{\perp}+\sigma^c_{\parallel}};\quad
\Delta L=\frac{L_{\perp}-L_{\parallel}}{L_{\perp}+L_{\parallel}},
\]
this gives 
\begin{equation}
\overline{\Delta n}=\frac{\Delta L+ \Delta\sigma }{1+\Delta\sigma\Delta L}.\label{eq:asymm-lum-1}
\end{equation}

In practice, the luminosity asymmetry depends only on the photon energy (and hence $W$). A preliminary fit is carried out for events binned only in $W$, and the values for $\Delta L$ fixed for the fits to individual $\{W,\cos\theta^{\star}_K\}$ bins.

By performing a fit to a mass spectrum such as Fig.\,\ref{fig:Missing_Mass} for the ${W,\cos\theta^{\star}_K}$ bin, a background fraction factor $\beta$ can be determined, which represents the ratio of the background cross section to the average of the combined cross sections in each setting:
\[
\beta=\frac{\sigma^{b}}{\frac{1}{2} \left( \sigma_{\perp}^{c}+\sigma_{\parallel}^{c} \right)}.
\]
This allows us to write 
\begin{equation}
\Delta \sigma =\left(1-\beta\right)\frac{\sigma_{\perp}^{s}-\sigma_{\parallel}^{s}}{\sigma_{\perp}^{s}+\sigma_{\parallel}^{s}},\label{eq:asymm-with-bkg}
\end{equation}
which can be connected with the expressions in equation (\ref{eq:f-and-g}).

One final point is that since each event is treated individually, provided that an independent estimate of the photon polarization can be made for that event, we do not need to worry about any difference in photon polarization in each setting. So for an event $i$ equation (\ref{eq:asymm-with-bkg}) becomes
\begin{equation}
\Delta \sigma=\left(1-\beta\right)\frac{P^{\gamma}_ig_i}{f_i},\label{eq:asyms-with-bkg-1}
\end{equation}
and plugging this into \ref{eq:asymm-lum-1} the final estimator is
\begin{equation}
\hat{a}_i=\frac{f_i \Delta L + (1-\beta) P^{\gamma}_i g_i }
{f_i + (1-\beta) P^{\gamma}_i g_i \Delta L}.
\label{eq:estimator}
\end{equation}
For each event measured $e_i$, the likelihood 
\[
{\cal P}_{i}\left(e_{i}\mid \xi, {\cal O},\lambda\right) = 
\frac{1}{2}\left(1+\hat{a}_i\left( \xi_i, {\cal O},\lambda \right)\right)
\]
is calculated. For the extraction of new observables, we use independently measured values of recoil polarization $P=p$ with uncertainties $\pm\delta p$
from interpolations of previous data \cite{mccracken_differential_2010,dey_differential_2010} as inputs. A  Normal probability density is then multiplied into the event likelihood:
\begin{equation}
{\cal P}_{i}\left(e_{i}\mid\xi_i, {\cal O},\lambda\right)\rightarrow{\cal P}_{i}\left(e_{i}\mid\xi_i, {\cal O},\lambda\right){\cal N}\left(P\mid\mu=p,\sigma=\delta p\right),\label{eq:likelihood-total}
\end{equation}
so that some variation in the value of $P$ is allowed in the likelihood
fitting of the asymmetry, but in a more constrained fashion. 

The total likelihood for all events in the $\left\{ W,\cos\theta^{\star}_K\right\} $ bin 
\begin{equation}
{\cal P}\left(\left\{ e_{i}\right\} \mid{\cal O},\lambda\right)=\prod_{i}{\cal P}_{i}\left(e_{i}\mid\xi_i, {\cal O},\lambda\right)\label{eq:likelihood-set}
\end{equation}
is maximized by varying the values of the observables ${\cal O}$.

The likelihood function is actually the probability of the data given
the parameters, whereas what we really want is the probability of
the parameters, given the data. This is given by the posterior probability
\begin{equation}
{\cal P}\left({\cal O}\mid\left\{ e_{i}\right\} \right)\propto{\cal P}\left(\left\{ e_{i}\right\} \mid{\cal O}\right){\cal P}\left({\cal O}\right),\label{eq:post}
\end{equation}
where we do not care about the normalizing constant, since the function
is to be maximized. So at the time of evaluating the likelihood, the
bounds $\left[-1,+1\right]$ are encoded into a prior probability
function ${\cal P}\left({\cal O}\right)$, since the support for values of the observables only exists in this region. This means our fit will
yield a maximum posterior probability estimate of the observables.

% === Bibliography  =======================================

\bibliographystyle{h-physrev}
\bibliography{g8_KY_references}% Produces the bibliography via BibTeX.

% === End of document =====================================

\end{document}